\documentclass[aps,prc,12pt,nofootinbib,superscriptaddress,showkeys,english]{revtex4-2}
\usepackage[english]{babel}
\usepackage{graphicx}
\usepackage{color}
\usepackage{amsmath}
\usepackage{slashed}
\usepackage{amssymb}
\usepackage{multirow}
\usepackage{color}
\usepackage{appendix}
\usepackage{enumitem}
\begin{document}
\title{Nonlinear electrodynamics for the vacuum of Dirac materials. Photon magnetic properties and radiation pressures}\author{A. W. Romero Jorge}
\email{adrian@icimaf.cu/aromerojorge@physik.uni-frankfurt.de}
\affiliation{Instituto de Cibern\'{e}tica, Matem\'{a}tica y F\'{\i}sica (ICIMAF), \\
            		Calle E esq a 15 Vedado 10400 La Habana Cuba}
\affiliation{Frankfurt Institute for Advanced Studies, Ruth Moufang Str. 1, 60438 Frankfurt, Germany}
		
\affiliation{Helmholtz Research Academy Hessen for FAIR (HFHF), GSI Helmholtz	Center for Heavy Ion Physics. Campus Frankfurt, 60438 Frankfurt, Germany}

\affiliation{Institut f\"ur Theoretische Physik, Johann Wolfgang Goethe	University, Max-von-Laue-Str. 1, 60438 Frankfurt, Germany}

	\author{A. P\'erez Mart\'inez}
 \email{aurora@icimaf.cu}
 \affiliation {Instituto de Cibern\'{e}tica, Matem\'{a}tica y F\'{\i}sica (ICIMAF), \\
		Calle E esq a 15 Vedado 10400 La Habana Cuba}
	\author{E. Rodr\'iguez Querts}
	\email{elizabeth@icimaf.cu}
 \affiliation{Instituto de Cibern\'{e}tica, Matem\'{a}tica y F\'{\i}sica (ICIMAF), \\
		Calle E esq a 15 Vedado 10400 La Habana Cuba}
\date{\today}
 % \title{Nonlinear electrodynamics for the vacuum of Dirac materials. Magnetic properties and anisotropic pressures}

\begin{abstract}
We investigate the magnetic properties of photons propagating through Dirac materials in a magnetic field, considering both vacuum and medium contributions. 
%We investigate the magnetic properties of Dirac materials in the presence of a magnetic field, considering both the medium and vacuum contributions. 
Photon propagation properties are obtained through a second-order expansion of non-linear Euler-Heisenberg electrodynamics at finite density and temperature considering Dirac material parameters
%($\alpha_D$, $\Delta$,
(Dirac fine structure constant, band gap
and Fermi velocity). Total magnetization (including electrons and photon contributions) and photon-effective magnetic moment are computed.  Observables such as photon energy density, radiation pressure, and Poynting vector are obtained by an average of components of the energy-momentum tensor. All quantities are expressed in terms of Lagrangian derivatives. Those related to the vacuum are valid for any value of the external magnetic field, and both the weak and strong field limits are recovered.  We discuss some ideas of experiments that may contribute to testing in Dirac materials the phenomenology of the strong magnetic field in the quantum electrodynamic vacuum and how non-linear corrections on the magnetization, radiation pressure, and birefringence,  are amplified up to $10^3$ times QED corrections.

\end{abstract}
\maketitle
\section{Introduction}

Maxwell’s theory predicts that plane-polarized light propagates in a vacuum, or empty space, at the speed of light 
$c$. Quantum Electrodynamics (QED), however, presents a different picture of the vacuum. According to this theory, the vacuum is not empty but rather a “sea” of virtual electron-positron pairs, with their negative energy states occupied and positive energy states unoccupied. This model explains how, in the presence of an external magnetic field, the speed of light remains $c$ when propagating parallel to the magnetic field but is lower than $c$ when propagating perpendicular to it. In this case, light changes its plane of polarization, resulting in the wave splitting into two polarized modes that move at different speeds (and frequencies). This phenomenon, known as Cotton-Mouton birefringence, is a consequence of the interaction between the magnetic field and the virtual electron-positron pairs
\cite{Rizzo:2010di, Rizzo2014}.

Viewed through the lens of Quantum Field Theories, the vacuum is considered the ground state of these theories. QED, consists of virtual pairs of electron-positrons, while for Quantum Chromodynamics, it involves virtual quarks-antiquarks. 
The shift from Maxwell's concept of vacuum to QED's dynamic ``sea" of virtual particle pairs redefines its conceptual meaning and aims to understand its properties and complexity.

%Illuminated by the Quantum field theories, the vacuum concept becomes in the ground state of the theories, being for QED, the virtual pairs of electron-positrons, and for Quantum Chromodynamics, vacuum virtual quarks-antiquarks. In summary, the shift from Maxwell’s vacuum to Quantum Electrodynamics’ dynamic “sea” of virtual particle pairs redefines its conceptual meaning and
%trying to understand its properties %and complexity.

%{\bf alguna oracion que cierre la idea del vacio }
The vacuum of QED in the presence of a strong electromagnetic field behaves as a medium, and virtual pairs interact with magnetic fields. Thus, the properties of light propagating in a QED vacuum can be studied using nonlinear optic tools, based on Maxwell's theory with all the magnitudes described in terms of a medium electric permittivity and magnetic permeability depending on the non-linear of the external electric and magnetic field\footnote{Vacuum electric permittivity and magnetic permeability are related with the speed of light \cite{Battesti}}.
Besides birefringence\footnote{In the presence of a strong electric field, birefringence also appears and is called Kerr birefringence}, QED predicts other exotic properties such as the Casimir effect, vacuum instability (close to the critical fields,  $B_c=\frac{m^2c^2}{e\hbar}=4.41\times 10^{13} $ G and $E_c=\frac{m^2c^3}{e\hbar}=1.3\times 10^{18} $ V/m,  where electron-positron pair creation occurs\cite{HEKockel, Heisenberg:1936nmg, Schwinger}), anisotropy pressures, etc. \cite{Battesti2012hf}.
All these vacuum phenomena predicted 90 years ago are still awaiting experimental confirmation. 
Testing the vacuum properties of QED requires large magnetic and/or electric fields. So far, technological barriers limit the generation of such high fields in the laboratory. The magnetic fields achieved in laboratories do not exceed a few teslas (around $5  $ T$= 5\times 10^4 $ G) \cite{field5T,Battesti2012hf}.
%Technologically, it has significant limitations.
For instance, vacuum birefringence is a tiny effect, even though it manifests for any magnetic field strength. At weak field limit, birefringence depends on a quadratic ratio between the magnetic field and the parameter  $\xi$, $\Delta n_{QED}\sim  (B/\xi)^2$, where $\Delta n_{QED}$ defined the difference between the refractive indices of the transverse modes  modes\footnote{$\xi=\frac{8\alpha^2\hbar^3}{45 m^4 c^5}\sim\frac{8\alpha}{45B_c^2}$, with $\alpha$ the fine structure constant.}.
The Polarization of Vacuum with Laser (PVLAS) experiment \cite{Ejlli:2020yhk} was designed to confirm the QED prediction of vacuum birefringence.  The strength of the magnetic field of the experiment was $2.5 \times 10^4$ G, so   $\Delta n_{QED}=2.5 \times 10^{-23}$, turning into a challenge, its measurement.
The experiment was improved, and it obtained a bound close to the prediction $\Delta n_{PVLAS}\sim (12\pm 17) \times 10^{-23}$ for the same magnetic field \cite{Ejlli:2020yhk}.

But in the Universe, we can find entities like Neutron stars, whose magnetic fields are close to the critical field, and even more in the case of magnetars \cite{Olausen2014, Turolla2017tqt, Hard}. Short-time huge magnetic fields may also be found in heavy-ion colliders, which can reach the order of $ 10^{18} $  G \cite{Zhong2014}. In both scenarios, neutron stars and heavy-ion colliders have reported indirect vacuum magnetic birefringence  \cite {Turolla2017tqt, Hard}. The work \cite {Turolla2017tqt} claims the first evidence of vacuum birefringence by measuring the degree and angle of polarization of the photons coming from the pulsar RX J1856.5-4 and comparing them with the theoretical results, but the uncertainties in their measurements are significant enough to trust in this measurement fully.

These observations have boosted the search for the phenomenon directly based on other experimental setups. One plausible alternative is the design of experiments with scattering of the pulsating laser \cite{Tommasini:2009nh}. It is predictable that with the developed technology of lasers, soon,  the power of lasers will reach an intensity around \textbf{$I\sim 10^{28}$ } Wm$^{-2}$ \cite{laser1,laser2} that corresponds to an electric (magnetic) field close to the critical field,  which may allow detection easier the effect of the birefringence of vacuum and to test the pair’s creation from the vacuum.

Besides, another exciting and new alternative to probe vacuum properties would be experiments designed with Dirac materials due to their critical fields  $E_{c}=\frac{\Delta^2}{ev_F} \sim  10^3 $ V/cm and $B_{c}=\frac{\Delta^2}{ev_F^2}\sim 10^4 \, \text{G} =1$ T are accessible in laboratories\footnote{where $ \Delta $ and $v_F$ are the band gap or energy gap and the Fermi velocity of the material respectively.}. 
A very hopeful work \cite{Keser2021yxp} has studied the contribution to the magnetization of the vacuum for three different 3D Dirac materials and the possibility of testing in experiments.

Dirac materials, including Weyl metals, topological materials,  and graphene, are characterized by electrons that behave as relativistic fermions with a linear dispersion relation. This unique property gives these materials distinctive optical, magnetic, and transport properties. The pioneer theoretical work of Dirac materials is \cite{teo}, but the boom of these materials only appeared with the fantastic finding of graphene by Novoselov \cite{Novoselov}. This experiment led to a boom of experimental and theoretical work to look for other materials or properties, and consequently, to technological applications.

The discovery of Dirac materials has been gratifying news for the Quantum Field Theory because it has opened the window to test in top-table experiments, theories, and phenomena requiring a large scale of energies and expensive experiments in particle accelerators \cite{toptableDirac}.
Consequently, the tools of Quantum Field theory have been extended to condensed matter physics. In line with this direction, the investigation of photon propagation parallel to an external magnetic field in electron-positron plasma \cite{Rojas82, Rojas79, Rojas87} was expanded to include graphene-like systems. This extension encompassed the study of the Faraday and Quantum Hall effects \cite{Richardpapiyyo, PerezMartinez2011eq, PhysRevA.88.052126}. Leveraging previous calculations \cite{Rojas82, Rojas79, Rojas87} from one-loop QED perturbative studies of 3D electron-positron plasma in a magnetized medium, a ``dimensional reduction" to 2D was performed while also considering the specific properties of graphene-like systems.
In this paper, we proceed similarly extrapolating to Dirac materials physics,  the results of the study of photon propagation transverse to the external magnetic field. We have used the extended to finite temperature and density of the effective Euler-Heisenberg (EH) non-linear electrodynamics to study the properties of electron-hole plasma. In particular, we use an approximated version of the extended EH Lagrangian, which corresponds to an expansion up to second order on the photon field \cite{PerezGarcia2022kvz} in the limit of $\omega\ll 2 m c^2$, where $m$ is the electron mass.
 From this Lagrangian, we investigate the magnetic and dielectric properties of a photon traveling in the Dirac vacuum and the observables, energy densities and pressures, and Poynting vector resulting from the energy-momentum tensor.
We also study the magnetic properties of Dirac materials in the medium and compare them with those of photons, finding that they are the primary contributors to overall magnetization.
 
%Theoretically, the properties of graphene are essentially described by Dirac massless fermions (electrons) in two dimensions. These systems are relativistic in that the spectra of electrons and holes can be mimicked as two-dimensional relativistic chiral fermions. Dirac materials also include those considered 3D but have topological properties that could be described by one loop Quantum Field Theory (QFT).

%These fascinating materials have awakened much attention to their applications and the possibility of studying QED properties using top-table experiments.

%We studied previously for graphene \cite{PerezMartinez:2011eq}, Quantum Hall Efect{PhysRevA.88.052126} and Faraday Effect from a one loop corrections results of QED extrapolated to the typical features of these materials.

%In this paper, we study the magnetic properties of Dirac materials in the presence of an external and constant magnetic field in the z-direction, considering the interaction of the magnetic field with virtual electron-hole pairs of Dirac vacuum and consequently the low photons excitations.
%For three-dimensional (3D) Dirac insulators and semimetals, low excitation photons correspond to the interaction of the magnetic field with the “Dirac vacuum,” i.e., filled valence band.
%Following the idea of our previous works, in this paper, we will “translate” our previous study of magnetic and dielectric properties of magnetized electron-positron plasma to the Dirac materials language.

Our work introduces several novel aspects that deserve highlighting. Firstly, we extend the effective Euler-Heisenberg action to finite density and temperature to discuss the magnetic properties of 3D Dirac materials. Additionally, we examine the magnetization of photons propagating transverse to an external magnetic field. To achieve this, we expand the Euler-Heisenberg Lagrangian up to the second order on the photon field, considering non-linear contributions from both vacuum and medium. This approach allows us to derive the magnetization of the Dirac materials described by electrons (Dirac vacuum and medium) as a zero-order term of the expansion and the magnetization of photons interacting with the magnetic field via the vacuum or medium.
Secondly, we consider the photons propagating in a Dirac vacuum while accounting for arbitrary magnetic field values, which allows for the recovery of both weak and strong magnetic field limits.
Thirdly, we calculate the energy-momentum tensor and observables for photons propagating in the vacuum in terms of derivatives of the effective Lagrangian. This formulation makes it possible to extend our approach straightforwardly to other non-electrodynamics. Furthermore, the extension of this calculation to finite temperature and density is also straightforward.
 
Furthermore, our study is also essential for the interest that awakens the magnetic properties of Dirac materials (including the vacuum properties) and for eventually testing the properties of the strong field QED regime. As we have already commented, critical fields for the Dirac materials are experimentally reachable. Hence, these materials could imitate the birefringence, anisotropic pressures, and other exotic properties of the vacuum of QED at the regime of the strong magnetic field.

The paper is organized as follows: in
Sec. II, a brief overview of Dirac materials characteristics is illustrated.
In Sec. III, we extrapolate the extended effective Lagrangian of QED in the presence of electromagnetic fields to 3D Dirac materials.
%presents the second-order expansion in the photon field of the extended effective Euler-Heisenberg Lagrangian. 
%
%The extending effective Lagrangian corresponds to considering finite temperature and density and the possibility of evaluating thermodynamical potential in the presence of external and constant magnetic fields. 
%
Sec. IV, is devoted to studying the electron and photon magnetic properties in the presence of an external magnetic field. 
In Sec. V, we calculate the photon magnetic moment. 
The solution of the Maxwell equation and dispersion equation of photon propagating in the 3D Dirac vacuum is presented in Sec.  VI.  
In Sec. VII, the energy-momentum tensor is presented, and observables like energy density pressures and Poynting vector are obtained.
Finally, we conclude the work, and some calculations are presented in the appendixes. We used natural units in the paper, except for some cases for clarity.

\section{A brief overview of Dirac materials characteristics} \label{sec2}

%In the context of cutting-edge materials, Dirac materials have gained increasing interest.
The crystalline structure of Dirac materials is characterized by a
“Dirac points” or well, with a small band gap
%(explicit q es, breve) in their band structure, where the valence and conduction bands touch at a single point, known as the Dirac point (figure \ref{puntoDirac}).
%These Dirac points
around them, the electrons behave like relativistic particles with free electrons as  Dirac fermions exhibiting an extremely high Fermi speed, about $10^6$ m/s
 and  electrons could be described by a linear relationship according to
 $\epsilon \approx v_F p$ between the energy and momentum of its electrons in the conduction and valence band for low energies \cite{Keser2021yxp, Neves2023uxi}.
 Dirac materials with zero mass (gap) are known as Weyl fermions, and the very well-known is graphene \cite{Cayssol}.
This one, a highly conductive material with no energy gap in its Dirac points
$(\Delta=0)$, it is distinguished from conventional semiconductors by its absence of this gap
$(\Delta\ne 0)$.
 These energy gaps can be controlled by manipulating the width of the graphene ribbon and introducing defects or doping, among other factors. By employing these techniques, small energy gaps are observed, ranging approximately from 100 meV to 250 meV \cite{Sahu2017,derivadografeno3,derivadografeno4,derivadografeno1,derivadografeno2}. Our study considers a non-zero gap graphene energy with an energy gap of 100 meV \cite{grafeno100}.

Formally, graphene is not  a 2D Dirac semimetal actually,
with the top of the valence and bottom of the conduction band of a band
The insulator just touches before gapping up again after band inversion, resulting in a graphene-like semimetallic state.
Besides graphene and topological insulators with 2D Dirac surface states, there is a 3D semimetal with a Dirac point in the bulk 3D Brillouin
    zone linear dispersion in all three $k$ directions. The theoretical prediction and the experimental proof in Bi-based materials did not take long \cite{PhysRevLett.95.146802, PhysRevB.74.195312, PhysRevLett.98.106803, PhysRevLett.96.106802}.
These materials are also called (3D) Dirac semimetals.

Our study will focus on these materials by considering a simple model.
We know that the crystal structure of these materials is sufficiently complicated with various nodes,
%(graphene has six nodes, TaAs 12 pairs of nodes, BiSb 19 nodes, and PnSnTe can have up to 14 nodes),
but as  a first approximation, we consider the following points:
\begin{enumerate}[itemsep=0pt]
    \item All nodes are equal and %On the other hand, the value of the energy gap of each material varies depending on the node and properties such as temperature, but
    we assume a constant energy gap independent of all the parameters of the material.
    \item We neglect the contribution to magnetization due to material impurities. 
    \item We are assuming that the properties of the materials are stable for values of the magnetic/electric field above the critical field of each material.  
\end{enumerate}

%%%%%%

%As a first approach to the study of Dirac materials %for %the theoretical calculation of their properties,
%we will consider a simple \textit{model for Dirac materials}.We know that the crystal structure of these materials is sufficiently complicated with various nodes.
%(graphene has 6 nodes, TaAs has 12 pairs of nodes, BiSb has 19 nodes, and PnSnTe can have up to 14 nodes).On the other hand, the value of the energy gap of each material varies depending on the node and properties. Therefore; we will consider the following approximations: All nodes are identical, and the energy gap is independent of all material parameters, with a constant value.
%coming by experimental measurements or an average %of its values. Each material will be defined by its Fermi velocity and energy gap, allowing various materials with similar properties to exist.

With this model, we will proceed to study the propagation of photons, extrapolating the results of QED, to 3D Dirac materials. It means shifting  the values of the critical fields
 in Dirac materials are defined as \cite{Keser2021yxp, Neves2023uxi}
\begin{equation}\label{key}
E_{c}(\Delta, v_F)=v_F B_{c}(\Delta, v_F)%\left( \dfrac{\Delta}{\alpha_D \lambda_D^3}\right)^{1/2}=
=\dfrac{\Delta^2}{e v_F},
\end{equation}
where the speed of light has been replaced by the Fermi velocity of the material ($ c\rightarrow v_F\approx10^6 $ m/s),
the fine structure constant has been changed to an effective or Dirac fine structure constant ($ \alpha \rightarrow \alpha_D $) defined as $ \alpha_D= e^2/ v_F \sim 1-4$. %$ \lambda_D $ is the Dirac’s Compton wavelength defined by $ \lambda_D= v_F/\Delta $.
Here the band gap or energy gap of the material is defined as $ \Delta=m^* m v_F^2 $, where $ m^*  \sim 0.01-0.5$  is the effective mass of electrons and holes \cite{Keser2021yxp, Neves2023uxi}.
%Here, $e$ and $m$ are the electron charge and mass respectively.
 
Table I presents a comparison of QED and four materials that we use in our study: graphene, tantalum arsenide ($Ta_3 As_4$), bismuth antimony ($Bi_{1-x}Sb_{x} $), and lead tin tellurium ($Pb_{1-x}Sn_{x}Te$).
%These values are based on experimental measurements or an average of their energy gap values.
%
Also, we have listed the values of $3\pi/\alpha_D$ for each material, which are related to the
maximum allowed values of the magnetic field to get accurate corrections to phase velocity in the strong field limit $ v_p^2\approx 1- \dfrac{\alpha_D}{ 3\pi} b >0$, %as it is shown in the next section,
where $b=B_e/B_c$, $B_e$ is the external constant magnetic field and $b_{max}=3 \pi/\alpha_D $. This bound ensures the validity of one-loop approximation and the effective non-linear electrodynamics \cite{DittrichLibro,PerezGarcia2022kvz}, above this strength, we have to include two-loop corrections.

 \begin{center}
    \begin{tabular}{|c|c|c|c|c|c|}
	\hline
	& $ \Delta $(meV)  & $ \alpha_D/\alpha $ & $ B_c $ (T)  & $ E_c $ (V/cm) & $\; b_{max} \;$ \\
	\hline
	QED   & $ 10^{9} $ & 1 & $ 4.4 \times 10^{9}$ & $1.3 \times 10^{16}$ &  $1291$\\
	\hline
	$Pb_{1-x}Sn_{x}Te $ & $ 31.5 $ & 580 & $ 5.6 $ &$2.9 \times 10^{4}$&$2.22$ \\
	\hline
	$Bi_{1-x}Sb_{x} $& $ 7.75 $ & 188 & $ 0.036 $ &$5.7 \times 10^{2}$&$6.86$\\
	\hline
	$Ta_3 As_4$ & 21 & 357 & $ 0.95 $&$7.9 \times 10^{3}$& $3.61$\\
	\hline
	Graphene & 100 &  301 & $ 15.4 $ &$1.5 \times 10^{5}$& $4.28$ \\
 \hline
%	Tinene & 72 &  416 & $ 16.9 $ & $3.10$ \\
% \hline
\end{tabular}
\end{center}
Table I. Comparison of QED and four Dirac materials for different parameters like band gap $(\Delta)$, Dirac structure constant ($ \alpha_D $) over $\alpha$, critical magnetic field ($B_c$), critical electric field ($E_c$),  and bound of the dimensionless external magnetic field $b=B_e/B_c$ imposed to ensure the validity of non-linear electrodynamics. Some values of QED are listed as a reference and the Dirac materials:
$Pb_{1-x}Sn_{x}Te$ \cite{Plomo2012,Temperature2,Keser2021yxp,Neves2023uxi},
	$Bi_{1-x}Sb_{x} $ \cite{Bismuto2008,Temperature2,Neves2023uxi,Keser2021yxp},
$Ta_3 As_4$ \cite{Temperature1,Keser2021yxp,Neves2023uxi},
and graphene \cite{Novoselov,PerezMartinez2011eq,grafeno100}.
%Structure constant , $\alpha$,
%$\alpha_D/\alpha = v_F^{-1}$, Fermi velocity of %each material $v_F \sim 10^6 $ m/s, and bound of %the validity of one loop approximation to the %dimensionless magnetic field.

Each material will be defined only by its Fermi speed and energy gap so that several materials can have the same characteristics. We are referring, for example, to the graphene-type material that has its Fermi speed and energy gap given, but it can be another material with the same characteristics.
%, equation (\ref{vstronglimit}).
%%%%%%%%%%%%%%%%%%%%%%%%%%%%%%%%%%%%%%%%%%%%%%%%%%%%%%%%%%%%%%%%%%%%%%%%%%%%%%%%%%%%%%%%%%%%%%%%%%%%%%%%
\section{NLED for Dirac materials, Lagrangian }
In this section, we extrapolate the extended effective Lagrangian of QED in the presence of electromagnetic fields to 3D Dirac materials.
The extending effective Lagrangian corresponds to considering finite temperature and density and the possibility of evaluating thermodynamical potential in the presence of external and constant magnetic fields. 
Besides, we can study the propagation of photons in 3D materials by doing an expansion of the effective Lagrangian up to the second order of the photon fields. 
This treatment allows us to calculate the magnetization of the electrons (zero-order of the expansion) and photons (terms of the first and second order of the expansion) and consider both the vacuum and medium contribution. 

We start from the Lagrangian of QED where  fermions are coupled with an electromagnetic vector field $A_{\mu}$\footnote{We used Minkowski flat space with the convention   
$g_{\mu\nu}=\eta_{\mu\nu}=
\{+1,-1,-1,-1\}$},
%to the physics 
%QED Lagrangian  has the form %
\newcommand{\fsl}[1]{{\ooalign{\(#1\)\cr\hidewidth\(/\)\hidewidth\cr}}}
\begin{equation}
\mathcal{L}(\bar{\psi},\psi,A_{\mu})=\bar{\psi}(\fsl{{\partial}} -e\slashed{A}_{\mu} -m)\psi +\frac{1}{4} F_{\mu\nu}F^{\mu\nu},
\end{equation}
\noindent where $\bar{\psi}$, $\psi$ are the fermion spinors, $\fsl{\partial}=\gamma^{\nu}\partial_{\nu}$, $\slashed{A}_{\mu}=\gamma^{\mu}A_{\mu}$, $\gamma_{\mu}$ are the Dirac matrices, $A_{\mu}$ is the cuatri-potential vector of the electromagnetic field and $F_{\mu\nu}$ is the electromagnetic tensor.
We start from  the grand partition function defined  as $Z=Tr \{ e^{k_B T (H-\mu N)} \}, $
where $k_B$ is the Boltzmann constant, $T$ is the temperature, $H$  is the Hamiltonian, $\mu$ is the chemical potential, and $N$ is the number of particle density.
The partition function in the path integral language corresponds to
\begin{equation}
Z=\int {\mathcal D}\bar{\psi}{\mathcal D}\psi e^{ik_B T\int_0^1 d\tau\int d^3x  {\mathcal L}(A_{\mu},\psi)},    
\end{equation}
where we use Euclidean space-time with  $\tau=it$ a variable in the interval $0 $ to $k_B T$.

In one-loop approximation, the effective Lagrangian at finite temperature and density\footnote{The thermodynamical potential of the electron-hole plasma is $\Omega=\mathcal{L}_{eff}$} in the presence of an electromagnetic field reads as,
\begin{equation}
\mathcal{L}_{eff}=k_B T \, Tr\{\ln{Z} \}=ik_B T \, \ln{ \det{ G^{-1}(x,x^{\prime})}}, \label{OmegaT}
\end{equation}
 where $G^{-1}(x,x^{\prime})$ 
 is the inverse electron Green function in the presence of electromagnetic fields and determinant and logarithm are functional operations. 
 %$S_{eff}_{D}(A)=\langle ...\rangle=\int d^4x \mathcal{L$  after %renormalization \cite{ren, Schwinger} (see appendix \ref{apA}). 
%
% In the momentum space, the electron propagator has the form
%\begin{equation}\label{Green-F-L}
%G^{-1}_n(\overline{p})= %\overline{p}\cdot\gamma-m,
%\end{equation}
%with the notation ${\overline{p}}=%(ip^{4},0, \sqrt{2eB_e n},p^{3})$ over %the Landau numbers $n=0,1,2,...$ in %Euclidean space, $\gamma$ are the Gamma-%matrices.
%In the momentum space, the %thermodynamical potential has the form %\cite{JorgitoyHugo, AuroraPRD2015}
Doing the functional operations over the propagator in integral representation, following the steps described in   \cite{Hattori:2023egw, Elmfors:1994fw, DittrichLibro} from Eq. (\ref{OmegaT}) one  arrives to the general effective Lagrangian \cite{Hattori:2023egw, Elmfors:1994fw, DittrichLibro}
%
%We can express this by shifting to Dirac materials, like
 %where $m\rightarrow\Delta$ and $c\rightarrow v_F$
\begin{align}	&\mathcal{L}_{eff}(\tilde{a},\tilde{b},\mu,T)=-\mathcal{F}-\frac{1}{8 \pi^2}\int_{0}^{i\infty} \frac{d s}{s^{3}} e^{-i(m^2-i\epsilon) s } \nonumber\\
	&\times\left [ (e s)^{2} \tilde{a} \tilde{b} \coth (e \tilde{a} s) \cot(e \tilde{b} s)
	\right ]\left( 1+ 2\sum^{\infty}_{k=1} e^{-\frac{iek^2 \beta^2}{4}h(s,\tilde{a},\tilde{b})}\cosh( \mu \beta k)\right) ,
 \label{eqEHTfinita}
\end{align}
where $\beta=1/k_B T$, % $k$ denotes the sum over the Poisson summation,
$\tilde{a}=\left[\left(\mathcal{F}^{2}+\mathcal{G}^{2}\right)^{1 / 2}+\mathcal{F}\right]^{1 / 2}$,
$\tilde{b}=\left[\left(\mathcal{F}^{2}+\mathcal{G}^{2}\right)^{1 / 2}-\right.$ $\mathcal{F}]^{1 / 2}$,    $\mathcal{F}$ and $\mathcal{G}$ are the secular invariants derived from the gauge and Lorentz invariants of the generic electromagnetic fields $({\bf E, B})$ 
%$\footnote{
%$\mathcal{F}$ and $\mathcal{G}$ %are  CPT invariant, the %dependency on effective %Lagrangian on  $\mathcal{G}^2$  %ensures the invariance although %$\mathcal{G}$ violates P and T. %Here C=Charge, P=Parity and %T=Time.}

\begin{align}
\mathcal{F}&=\frac{1}{4}F^{\mu\nu}F_{\mu\nu}=\frac{1}{2}(-\epsilon_0 E_e^2+\frac{B_e^2}{\mu_0}),\\ \mathcal{G}&=\frac{1}{4}F^{\mu\nu}\tilde{F^{\mu\nu}}=\sqrt{\frac{\epsilon_0}{\mu_0}}\boldsymbol{(-E_e \cdot B_e)},
\end{align}
where $\mu_{0}$ and $\epsilon_{0}$ are electrical permittivity and magnetic permeability respectively\footnote{Along the paper, by simplicity, we take  $\mu_{0}=\epsilon_{0}=1$}. $E_{e,(i)}$ and $B_{e,(i)}$ are the components of electromagnetic tensors defined as $E_{e,(i)}=F_{0 i}$, $B_{e,(i)}=-\frac{1}{2} \epsilon_{i j k} F^{j k}$  with $i=1,2,3$, $\tilde{F}^{\mu\nu}=\epsilon^{\mu\nu\alpha\beta}F_{\alpha\beta}/2$ is the dual tensor, and $\epsilon^{\mu\nu\alpha}$ and  $\epsilon^{\mu\nu\alpha\beta}$ are the totally antisymmetric Levi-Civita tensors of rank 3 and 4, respectively.  

The first term of the effective Lagrangian is the unrenormalized non-linear EH Lagrangian, while the second one corresponds to the temperature and density correction. In a medium, Lorentz symmetry is broken, and the reference frame would be specified by the medium velocity $u_{\mu}$. Hence, the general effective Lagrangian apart of the electromagnetic field invariants $\tilde{a}$, $\tilde{b}$ includes $u_{\mu}=(1,0,0,0)$ by the covariant form of the electric field $E_e^{\mu}=F^{\mu\nu}u_{\nu}$,  $E^\mu_e=(0,\textbf{E}_e)$ so that $\varepsilon_{u}^2=|\textbf{E}_e|^2$.The function $h(s)$ that appears in the exponent of the second term is 
\begin{equation}
 h(s,\tilde{a},\tilde{b})= \tilde{a}\frac{\tilde{b}^2+\varepsilon_{u}^2}{\tilde{a}^2+\tilde{b}^2}\cot(e\tilde{a}s)+
\tilde{b}\frac{\tilde{a}^2-\varepsilon_{u}^2}{\tilde{a}^2+\tilde{b}^2}coth(e\tilde{b}s),
\end{equation}
and it relates the electromagnetic invariants with medium velocity (details of the extended effective Lagrangian could be seen in \cite{DittrichLibro, Hattori:2023egw, Elmfors:1994fw}).  
 
 The effective Lagrangian Eq (\ref{eqEHTfinita}) is reduced to constant external magnetic field taking $\tilde{b}\rightarrow 0$ and $\mathcal{G}=0$ ($E_e=0$)  leaving only the invariant $\mathcal{F}=B_e^{2}/2$. 
 
 In that case, the first term of the Lagrangian depends only on the magnetic field, while the second term additionally depends on temperature and chemical potential, 
\noindent with
\begin{align}
\mathcal{L}_{eff}(B_e)&=-\frac{1}{8\pi^2}\int_{\epsilon}^{\infty} \frac{ds}{s^3}\left( 2e^{-m^2s}(eB_es)\coth (eB_es) \right )\label{vacBpropertime},\\
&=\frac{eB_e}{4\pi^2} \int_{-\infty}^{\infty}  dp_3  \sum_{\sigma ,n} \vert E_{\sigma n} \vert,\label{vacBimaginarytime}
\end{align}
where $\mathcal{L}_{eff}(B_e)$  has to be renormalized
and with the Maxwell classical terms we get the Euler-Heisenberg effective non-linear Lagrangian \cite{HEKockel}.  Its expression is  
\begin{equation}
\mathcal{L}^R_{eff}(B_e)
=-\mathcal{F}-\frac{1}{8\pi^2}\int_{\epsilon}^{\infty} \frac{ds}{s^3}\left( 2e^{-m^2s}(eB_es)\coth (eB_es)-1-\frac{(eB_es)^2}{3}\right).\label{Lrenor}
\end{equation}

The $\mathcal{L}_{eff}(B_e,T,\mu)$ remains as \cite{DittrichLibro}
\begin{align}
\mathcal{L}_{eff}(B_e,\mu,T)&=-\frac{1}{8\pi^2}\int_{\epsilon}^{\infty} \frac{ds}{s^3}2e^{-m^2s}\left((eB_es)\coth (eB_es)\right )\left( \sum^{\infty}_{k=1} e^{-\frac{ek^2 \beta^2}{4s}} cosh(\mu \beta k)\right )\label{medBpropertime}\\
&=\frac{eB_e}{4\pi^2} \int_{-\infty}^{\infty}  dp_3 \sum_{\sigma ,n} \frac{1}{\beta} \ln(1+e^{-\beta\vert E_{\sigma n}  -\mu\vert})(1+e^{-\beta\vert E_{\sigma n} +\mu\vert}),
\label{medBimaginarytime2}
 \end{align}
where  $E_{\sigma n}=\sqrt{p_3^2+m^2+eB_e(2n+1+\sigma)}$,
%$E_{\sigma n}=\sqrt{p_3^2v_F^2+\Delta^2+eB_e %v_F(2n+1+\sigma)}$ is the spectrum of electrons in Dirac materials,
$\sigma$ is the spin and the number $k$ corresponds to the sum over the Matsubara frequencies.

We can analyze their contributions separately as 

 %The first term on the right-hand side of Eq (\ref{eqEHTfinita}) depends only on the magnetic field, while the second term depends on temperature and chemical potential,
 %enabling us to analyze both contributions as follows separately}
\begin{equation}\label{omega-2}
\mathcal{L}_{eff}(B_e,T,\mu)=\mathcal{L}^R_{eff}(B_e)+\mathcal{L}_{eff}(B_e,T,\mu).\end{equation}
We want to point out that Eq. (\ref{vacBimaginarytime}) and (\ref{medBimaginarytime2})  
show the equivalence in one loop of the effective Lagrangian \cite{Hattori:2023egw}  obtained by proper time Schwinger method Eq. (\ref{vacBpropertime}) and (\ref{medBpropertime}) respectively and the obtained using the functional approach in imaginary time formalism for electron propagator interacting with the magnetic field at finite temperature and density \cite{JorgitoyHugo}. (The appendix (\ref{equivalencia}) provides details on how to transition from one to the other). 
Although both representations are equivalent,  
depending on the specific calculation, one might be more advantageous than the other. For instance, the integral representation using the proper time method is simpler for regularizing the effective Lagrangian. Conversely, for calculating integrals at the limit of zero temperature and non-zero chemical potential, the partition function of electron-positrons 
%imaginary time method 
is more suitable and straightforward.
%Even when both methods are equivalent depending on what you want to calculate,  one or the other will be more convenient. For example, to regularize the effective Lagrangian, it is simpler to do using the proper time method, while to calculate integrals at the limit of zero temperature and non-zero chemical potential, the imaginary time method is more adequate and simple.

Usually, when one studies electron-positron plasma in an external magnetic field, the vacuum contribution is ignored because it is only relevant close to the critical field.
However,  for Dirac materials, the critical fields are reachable in laboratories, so studying the relevance and the presence of their contribution is very timely for backing up the QED vacuum properties.

Let's move to  Dirac material language. Without loss of generality \cite{Keser2021yxp,Neves2023uxi}, we  change  $m $ and $\alpha$ in the effective Lagrangian by $\Delta$ and $\alpha_D$ Eq. (\ref{eqEHTfinita})-(\ref{medBimaginarytime2}), i.e. from now on $\mathcal{L}_{eff}
\rightarrow\mathcal{L}_{D}(\alpha_D,\Delta)$ . 
%From now on we change the notation of effective Lagrangian $\mathcal{L}_{eff}$ by $\mathcal{L}_{D}$ to identify 3D Dirac materials effective Lagrangian.

Experiments on Dirac materials typically involve densities higher than temperatures of around $10^2$ meV, while the temperatures are approximately $10$ K ($<1$ meV) \cite{Temperature1, Temperature2}.
So, the  degenerate limit ($T\ll \mu$) is suitable,  holes are neglected and the integral over $p_3$ in Eq. (\ref{medBimaginarytime2}) yields the following expression for the effective Lagrangian \cite{Felipe:2002wt} 
\begin{align}
\mathcal{L}_D=  \frac{e\Delta^2}{4 \pi^2 v_F} \sum_{n=0}^{n_{\mu}} a_n b \int \mid E_{\sigma n}^{\prime}-\mu^{\prime} \mid \theta(E_{\sigma n}^{\prime}-\mu^{\prime})dp_3^{\prime},\\
= \frac{e\Delta^2}{4 \pi^2 v_F} \sum_{n=0}^{n_{\mu}} a_n b\left (\mu^{\prime} p_f^{\prime}-(1+2nb  )ln\left (\frac{\mu+p_f}{\Delta_n }\right )\right ), \label{OmegaBT0}
\end{align}
 where $a_n=2-\delta_{0n}$, $b=B_e/B_c$, $dp_3^{\prime}=dp_3/\Delta$,  $E_{\sigma n}^{\prime}=\sqrt{p_3^2/\Delta^2+1+(2n+1+\sigma)b}$, 
$\mu^{\prime}=\mu/\Delta$, 
$p^{\prime}_f=1/v_F \sqrt{\mu^{\prime 2}-\Delta_n^{\prime 2}}$,
$\Delta_n^{\prime 2} =1+2nb$,
$p_f=1/v_F \sqrt{\mu^2-\Delta_n^2}$ is the Fermi momentum, $\Delta_n^2=\Delta^2+2eBnv_F$, and
 $n_{\mu}$ is the maximum number of Landau levels taken from the integer part of $n_{\mu}=\frac{\mu^2-\Delta^2}{2eB_ev_F}=\frac{\mu^2-1}{2b}$.  
%From the thermodynamical potential, it is possible to calculate the magnetization of the system as $\mathcal{M}^{medium}=-\frac{\partial\Omega}{\partial B}$ and the results yield

%In the next section, we look for an effective theory to study this problem using the extended effective EH Lagrangian Eq. (\ref{eqEHTfinita}). 

\subsection{Photon propagation in the presence of a constant magnetic field}
 We aim to investigate the properties of photons propagating perpendicular to an external and constant magnetic field along the $x_3$-direction in a 3D Dirac material. To undertake this study, we employ non-linear electrodynamics (non-perturbative QED), beginning with the extension of Euler-Heisenberg's effective Lagrangian to finite temperature and density regimes in the infrared limit ($\omega << 2m^2$).

Therefore, we start from  effective Lagrangian Eq. (\ref{eqEHTfinita}) making the prescription for  the vector potential \cite{PerezGarcia2022kvz}: $A_{\mu}=A_{\mu}^{ext}+a_{\mu}$, where $a_{\mu}$ represents the four-potential of the photon and $A^{ext}_{\mu}$ is the potential associated with the external and constant magnet  ic field $\mathbf{B}_e$ and the corresponding electromagnetic field tensor is
$ \mathcal{F}_{\mu \nu}= f_{\mu \nu} + \mathcal{F}^{B}_{\mu\nu}$  where
\begin{align}
%\mathcal{F}_{\mu\nu}=f_{\mu\nu} + \mathcal{F}^{B}_{\mu\nu}, \quad \quad
f_{\mu\nu}=\partial^{\mu}a^{\nu}-\partial^{\nu}a^{\mu},\quad \quad
\mathcal{F}^{B}_{\mu\nu}=\partial^{\mu}A^{ext,\nu}-\partial^{\nu}A^{ext,\mu},
\end{align}
 with
%$%=(0, B_e^3)$
%$ and
$\mathbf {B=B_e+B_w}$ and $\mathbf{E=E_w}$, where $B_w$ and $E_w$  are the wave magnetic and electric field respectively. The invariants become in
\begin{align}
\mathcal{F}&=\frac{1}{4}F^{\mu\nu}F_{\mu\nu}+
\frac{1}{4}f^{\mu\nu}f_{\mu\nu}=\frac{1}{2}(-E_w^2+B_w^2) +  \frac{B_e^2}{2} ,\\
\mathcal{G}&=\frac{1}{4}F^{\mu\nu}\tilde{F^{\mu\nu}}+ \frac{1}{4}f^{\mu\nu}\tilde{f^{\mu\nu}}+\frac{1}{4}f^{\mu\nu}\tilde{F^{\mu\nu}}+\frac{1}{4}F^{\mu\nu}\tilde{f^{\mu\nu}}=\boldsymbol{-E_w \cdot B_e}.\label{waveinvariant}
\end{align}
%The prescription allows to do an  expansion of the thermodynamical potential   (\ref{eqEHTfinita}) in terms of the invariants  (\ref{waveinvariant}) up to the second order of the photon field. \textbf{Details of  NLED for QED  can be seen in \cite{PerezGarcia2022kvz}}.
As the photon propagates in constant magnetic field,  $E\rightarrow 0$, i.e. and $\tilde{b}\rightarrow0$ and  $h(s)=1/s$. We expand up to the second order of the photon fields the extended effective Euler Heisenberg Lagrangian \cite{PerezGarcia2022kvz, DittrichLibro}.     

%Writing the effective  termodynamical potential in terms of the photon fields $({\bf E}_w, {\bf B}_w)$ 
The explicit expression of the expansion has the form 
\begin{align}
\mathcal{L}_D(B_e,E_w,B_w,T,\mu)=\mathcal{L}_D^{(e-B_e)}(B_e,T,\mu)\mid_{E_w=B_w=0} + \mathcal{L}_D^{(ph)}(B_e,E_w,B_w,T,\mu),\label{expansionL}
\end{align}
where $\mathcal{L}_D^{(e-B_e)}(B_e,T,\mu)\mid_{E_w=B_w=0}$ is the zero order of the expansion,  and it corresponds to the effective Euler Heisenberg Lagrangian for electrons in the presence of constant magnetic field written in the previous heading as Eq. (\ref{vacBimaginarytime}) and (\ref{medBimaginarytime2}), $\mathcal{L}^{(e-B_e)}_D(B_e,T,\mu)\mid_{E_w=B_w=0}=\mathcal{L}^R_D(B_e)+\mathcal{L}_D(B_e,T,\mu)$,  
%(\ref{OmegaT}) 
%\begin{equation}
%&\mathcal{L}^{e-B}_D(B_e)=\mathcal{L}_D(B_e)\left ( %1+\sum^{\infty}_{n=1} e^{-\frac{en^2}{4sT^4}} %cosh(\frac{en\mu}{T})\right ),\nonumber\\
%\end{equation}
while 
$\mathcal{L}_D^{(ph)}(B_e,E_w,B_w,T,\mu)$ is a quadratic non-linear effective Lagrangian and it  contents interaction of photons with the magnetic field via virtual pairs (vacuum) and through electrons (medium) and, in the considered approximation, looks like
\begin{align}
\mathcal{L}_D^{(ph)}=\frac{1}{4}(1-{\mathcal{L}}^{f,D}_{\mathcal F}) f^{\mu\nu}f_{\mu\nu}
+\frac{\mathcal{L}^{f,D}_{\mathcal{FF}}}{8} (f^{\mu\nu}F^B_{\mu\nu})^2
+\frac{\mathcal{L}^{f,D}_{\mathcal{GG}}}{8} (f^{\mu\nu}\tilde{F}^B_{\mu\nu})^2)\label{NLEDT},
\end{align}
\noindent where the coefficients of the expansion are ${\mathcal{L}}^{f,D}_{\mathcal F}={\mathcal{L}}^{D}_{\mathcal{ F}}+{\mathcal L}^{(\mu,T),D}_{\mathcal F}$,   ${\mathcal{L}}^{D}_{\mathcal{ F}}$ is the renormalized EH effetive action Eq.  
${\mathcal{L}}^{f,D}_{\mathcal {FF}}={\mathcal{L}}^{D}_{\mathcal FF}+{\mathcal L}^{(\mu,T),D}_{\mathcal{FF}}$,
$\mathcal{L}^{f,D}_{\mathcal {GG}}={\mathcal{L}}^{D}_{\mathcal{GG}}+\mathcal{L}^{(\mu,T),D}_{\mathcal {GG}}$.

The presence of finite temperature and density, which characterize the medium, appear as temperature and density-dependent expansion coefficients in addition to those that depend solely on the magnetic field, such as $\mathcal{L}^D_{\mathcal{F}}$, $\mathcal{L}^D_{\mathcal{FF}}$, and $\mathcal{L}^D_{\mathcal{GG}}$. These latter are regularized integrals dependent on arbitrary values of the external magnetic field 
%while the former are integrals that also depend on temperature and chemical potential 
\cite{PerezGarcia2022kvz, DittrichLibro, Karbstein2013, Karbstein2015, Karbstein2021}. They represent quantum corrections up to second order to the classical Lagrangian photon field $f^{\mu\nu}f_{\mu\nu}$,  proportional to  $\alpha_D$. We can rewrite Eq. (\ref{NLEDT}) by separating the vacuum from the medium  photon interaction, $\mathcal{L}_D^{(ph)}=\mathcal{L}_D^{(ph-B_e)} +\mathcal{L}_D^{(ph-B_e-e)}$ where 
\begin{align}
\mathcal{L}^{(ph-B_e)}_D&=\frac{1}{4}(1-{\mathcal{L}}^D_{\mathcal F}) f^{\mu\nu}f_{\mu\nu}
+\frac{\mathcal{L}^D_{\mathcal{FF}}}{8} (f^{\mu\nu}F^B_{\mu\nu})^2
+\frac{\mathcal{L}^D_{\mathcal{GG}}}{8} \label{ELagvac}(f^{\mu\nu}\tilde{F}^B_{\mu\nu})^2, \\
\mathcal{L}^{(ph-B_e-e)}_D&=\frac{1}{4}{\mathcal L}^{(\mu,T),D}_{\mathcal F} f^{\mu\nu}f_{\mu\nu}
+\frac{\mathcal{L}^{(\mu,T),D}_{\mathcal{FF}}}{8} (f^{\mu\nu}F^B_{\mu\nu})^2
+\frac{\mathcal{L}^{(\mu,T),D}_{\mathcal{GG}}}{8} (f^{\mu\nu}\tilde{F}^B_{\mu\nu})^2.\label{NLEDT2}
\end{align}
The integrals concerning Lagrangian derivatives are outlined in the appendix (\ref{LagrangianderivativesatT0mu0}) and (\ref{DerivativesLagrangianT0mune0}). For a detailed explanation of the calculations, see \cite{DittrichLibro, Hattori:2023egw}.

Let us note that all the Lagrangians Eqs. (\ref{eqEHTfinita}), (\ref{NLEDT}) and (\ref{NLEDT2}) preserve the parity conservation, because there are no odd powers of the invariant $\mathcal{G}$ in these expressions,  since
both coefficients $\mathcal{L}_{\mathcal{FG}}=\mathcal{L}_{\mathcal{G}}=0$ .  
\section{Magnetic properties of photon propagation in Dirac materials. Magnetization}\label{sec:mag}

Let us start from Eq. (\ref{expansionL}) to study the total magnetization resulting from electrons and photons interacting with the magnetic field via vacuum (virtual pairs) and medium (electrons). 
%\begin{equation}
%\Omega_D^{\rm (ph-B_e)}\equiv
%\langle\frac{(1-%\mathcal{L}_{\mathcal{F}})}{2}(E_w^2-B_w^2)+
%\frac{\mathcal{L}_{\mathcal{FF}}}{2}({\bf B_e} \cdot{\bf B}_w)^2+ \frac{\mathcal{L}_{\mathcal{GG}}}{2}({\bf B_e}\cdot {\bf E}_w)^2\rangle,
%\label{2orderLagrangianE2B}
%\end{equation}
%\textbf{where $\langle... \rangle \equiv \int d^3 x$  denotes the average over coordinates of the photon fields.}
%The medium magnetization $\Omega^D_{e-B_e}$ is the thermodynamical potential of the electrons Dirac material plasma  \cite{PerezMartinez2011eq}.
This is obtained by 
$\mathcal{M}^{(T)}(B_e)=-\frac{\partial\mathcal{L}_D}{\partial {\bf B}_e}$  \cite{PerezMartinez2011eq}
and it has the form 
\begin{align}
\mathcal{M}^{(T)}&=\mathcal{M}^{(e)}(B_e,T,\mu)+{\mathcal M}^{(ph)}(B_e,T,\mu),\\
{\mathcal M}^{(e)}(B_e,T,\mu)&={\mathcal M}^{(B_e)}(B_e)+{\mathcal M}^{(e-B_e)}(B_e,T,\mu),\\
{\mathcal M}^{(ph)}(B_e,T,\mu)&={\mathcal M}^{(ph-B_e)}(B_e)+{\mathcal M}^{(ph-B_e-e)}(B_e,T,\mu).
\end{align}
 In the  degenerate gas limit (at finite density and zero temperature), the Dirac material magnetization (due to electrons)  can  be  calculated from Eq. (\ref{OmegaBT0})  and it reads as
\begin{equation}
\mathcal{M}^{(e-B_e)}(B_e,\mu)= \mathcal{M}_0 \sum_{n=0}^{n_{\mu}} \frac{3 a_n}{2} \left (\mu^{\prime}p_F^{\prime}- \left(1+ 4n b\right) ln\left (\frac{\mu+p_f}{\Delta_n }\right )\right ),
\end{equation}
where $\mathcal{M}_0$=$ e \Delta^2/(6\pi^2v_F)$ \cite{Felipe:2002wt}.
 %, $a_n=2-\delta_{0n}$,
%$\mu^{\prime}=\mu/\Delta$, 
%$p^{\prime}_f=1/v_F \sqrt{\mu^{\prime 2}-1-2nB_e/B_c}$. 
The corresponding magnetization of the vacuum is obtained from Eq. (\ref{vacBpropertime}), which arises from the interaction of virtual pairs with the magnetic field. 
The integral can be solved for weak (WF) and strong field (SF) limits, yielding
\begin{equation}
\mathcal{M}^{(B_e) SF}(B_e)=\mathcal{M}_0 \;  b(\ln(b)+1), \quad 
\mathcal{M}^{(B_e) WF}(B_e)=\mathcal{M}_0 \; b^3. \label{Mag_Be}
\end{equation}
%
%\textbf{OJO, esto es cierto pero tenemos q ajustar las unidades para que la del medio de los materiales nos de del mismo orden q esta 
This vacuum magnetization $\mathcal{M}^{(B_e)}(B_e)$ agrees with \cite{Keser2021yxp} for $E_e=0$. 

Photon magnetization due to the photon interaction with the vacuum is
\begin{align}
{\mathcal M}^{(ph-B_e)}(B_e)&=\langle\mathcal{L}_{\mathcal{GG}}^D(B_e) ( {\bf E}_w \cdot {\bf B}_e) {\bf E}_{w}  +\mathcal{L}_{\mathcal{FF}}^D(B_e) ({\bf B}_w \cdot {\bf B}_e ) {\bf B}_{w}\rangle,
\end{align}
and with medium 
\begin{align}
{\mathcal M}^{(ph-B_e-e)}(B_e,T,\mu)&=\langle\mathcal{L}^{(\mu,T),D}_{\mathcal{GG}}(B_e) ( {\bf E}_w \cdot {\bf B}_e) {\bf E}_{w}  +\mathcal{L}^{(\mu,T),D}_{\mathcal{FF}}(B_e) ({\bf B}_w \cdot {\bf B}_e ) {\bf B}_{w}\rangle\label{magnetizationf1},
%=&\langle\mathcal{L}^T_{\mathcal{GG}}(B_e) ( {\bf E}_w \cdot {\bf B}_e) {\bf %E}_{w}  +\mathcal{L}^T_{\mathcal{FF}}(B_e) ({\bf B}_w \cdot {\bf B}_e ) {\bf %B}_{w}\rangle,
%({\bf B}_e +{\bf B}_w).
\end{align}
Note that $\langle ...\rangle$ indicates the average over space-time, which applies only to the photon field since it is the only quantity dependent on space-time, as a plane wave.

The magnetization of photons exhibits the same structure in both vacuum and medium: a linear dependence on the magnetic field and a quadratic dependence on the photon fields $E_w^2$ and $B_w^2$. The distinction between vacuum and medium arises from the expressions of the second Lagrangian derivatives with respect to the external magnetic field. These derivatives for the vacuum correspond to finite integrals that are functions of the magnetic field: $\mathcal{L}^D_{\mathcal{FF}}$ and $\mathcal{L}^D_{\mathcal{GG}}$, while $\mathcal{L}_{\mathcal{FF}}^{(\mu,T),D}$ and $\mathcal{L}_{\mathcal{GG}}^{(\mu,T),D}$ depend on temperature, density, and magnetic field (see the appendix (\ref{LagrangianderivativesatT0mu0}) and (\ref{DerivativesLagrangianT0mune0}) for their expressions). 

% Comentado pq see repite, ya arriba see dijo lo del limte degenerado
%In our study we are interested in the degenerate limit ($T\ll\mu$).

%for the scalars $\mathcal{L}^{\mu}_{\mathcal{FF}}$ and  $\mathcal{L}^{\mu}_{\mathcal{GG}}$. 
The explicit expressions are in the Appendix (\ref{DerivativesLagrangianT0mune0}).  
For each polarization mode, $i=2,3$ the photon magnetization in the medium could be separated as
%Eq. (\ref{magnetizationf1}), %(\ref{LGGLFFM}) and  (\ref{LGGLFFM2})
\begin{equation}
\mathcal{M}^{(2)\,\,(ph-B_e-e)}_{}=\mathcal{L_{GG}}^{(\mu),D}B_e \langle E_w^2\rangle, \quad\quad \mathcal{M}^{(3)\,\,(ph-B_e-e)}=\mathcal{L_{FF}}^{(\mu),D}B_e \langle B_w^2\rangle, \label{mag232}
\end{equation}
while  for  vacuum, 
%Fig. (\ref{fig:modes}) as
  %happens, for instance, for the presence of external magnetic and electric fields with an electric field lower than the magnetic field $E_e\ll B_e$.

\begin{equation}
\mathcal{M}^{(2)(ph-B_e)}=\mathcal{L}^D_{\mathcal{GG}}B_e \left\langle E_w^2\right\rangle,  \quad \quad \mathcal{M}^{(3)(ph-B_e)}=\mathcal{L}^D_{\mathcal{FF}}B_e \left\langle B_w^2\right\rangle.  \label{mag23}
\end{equation}
%\textbf{where we will take $E_\omega=v_F %B_\omega$ for later calculations. }
Eq. (\ref{mag23})  reproduces the photon vacuum magnetization in the weak and strong magnetic field limits \cite{Elizabeth, Villalba-Chavez:2012pmx}  as follows    
\begin{align}
{\mathcal M}^{(2),WF}_{(ph-B_e)}= \frac{7\xi_D}{2} B_e \langle E_w^2\rangle\pmb{\hat{x_3}}, \quad \quad
{\mathcal M}^{(3),WF}_{(ph-B_e)}=2\xi_D B_e\langle B_w^2\rangle\pmb{\hat{x_3}},\\
{\mathcal M}^{(2), SF}_{(ph-B_e)} =\frac{\alpha_D}{3\pi }\frac{\langle E_w^2\rangle}{B_c} \pmb{\hat{x_3}}, \quad \quad
{\mathcal M}^{(3),SF}_{(ph-B_e)} =\frac{\alpha_D}{3\pi}\frac{\langle B_w^2 \rangle}{B_e}\pmb{\hat{x_3}}.
\label{magnetizacion}
\end{align}

\begin{figure}[h!]
\centering
\includegraphics[width=0.7\linewidth]{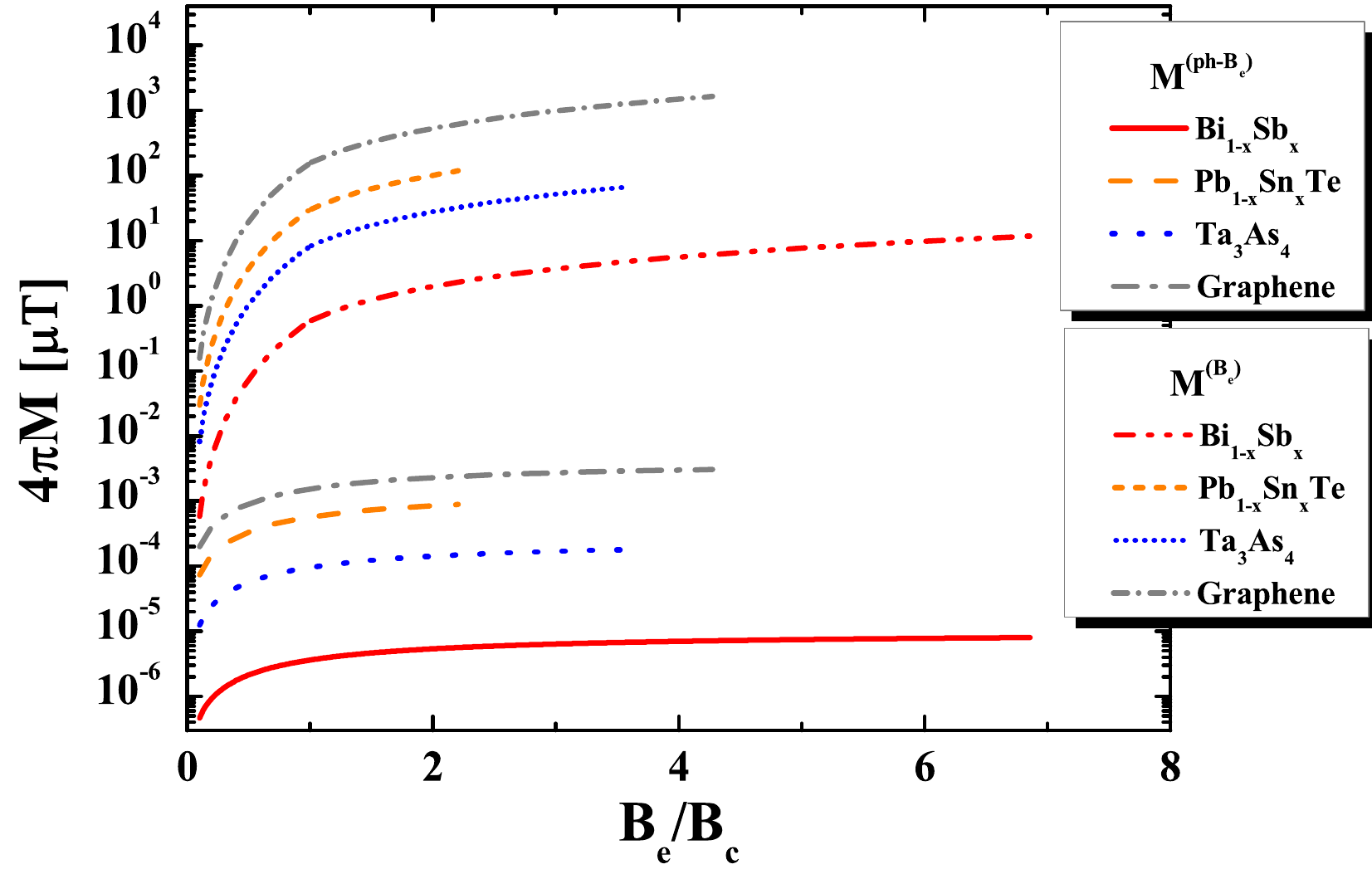}
\caption{Vacuum magnetization $\mathcal{M}^{(B_e)}$ due to the interaction of virtual pairs with the magnetic field,   (represented by the bottom four curves), and photon magnetization $\mathcal{M}^{(ph-B_e)}$ resulting from their interaction with the magnetic field through virtual pairs,  (represented by the top four curves), are shown for mode $i=2$. The magnetization is plotted as a function of the dimensionless magnetic field $b=B_e/B_c$ for each Dirac material:  $Bi_{1-x}Sb_{x} $, %(solid and dashed-dot-dot red)
 $Pb_{1-x}Sn_{x}Te$, %(dashed and short dashed orange)
  $Ta_3 As_4$, %(dotted and short dotted blue)
 and graphene. %(dashed dot and dashed-dot-dash gray)
  The value of $E_w$ is normalized using the electrical critical field, $E_\omega/E_c=0.01$, and $E_\omega = v_F B_\omega$. The curves are plotted up to the maximum value of the magnetic field according to the validity in the one-loop approximation, $b_{max}$ in Table I. }
\label{Mag_Vacuum}
\end{figure}

 Fig. (\ref{Mag_Vacuum}) shows a comparison between the vacuum magnetization $\mathcal{M}^{(B_e)}$ (Eq. (\ref{Mag_Be})) and the photon magnetization $\mathcal{M}^{(ph-B_e)}$ (Eq. (\ref{mag23})) as a function of the external magnetic field for the  Dirac materials of Table I. We used $E_\omega/E_c=0.01$ and $E_w=v_F B_w$ to satisfy the expansion requirement ($E_w \ll E_c$ and $B_w \ll B_c$). It is observed that the photon magnetization is always lower than the electron vacuum magnetization.
While the photon magnetization varies between $10^{-6}$ $\mu$T and $10^{-3}$ $\mu$T for different materials, the contribution of electron vacuum magnetization is 3 to 6 orders of magnitude greater. In both cases, the magnetization increases with the strength of the external magnetic field and tends to remain constant when the magnetic field reaches the corresponding critical field for each material.

The contribution of the vacuum to the total magnetization is tiny at weak magnetic field scales; however, at higher magnetic field values, this contribution increases and becomes nearly constant. The photon magnetization  $\mathcal{M}^{(ph-B_e)}$ of $Bi_{1-x}Sb_x$ for $B_e/B_c=0.1$ is approximately on the order of $10^{-14}$ T, whereas for $Pb_{1-x}Sn_xTe$, it is on the order of $10^{-12}$ T for the same external magnetic field value.

\begin{figure}[h!]
\centering
  \includegraphics[width=0.49\linewidth]{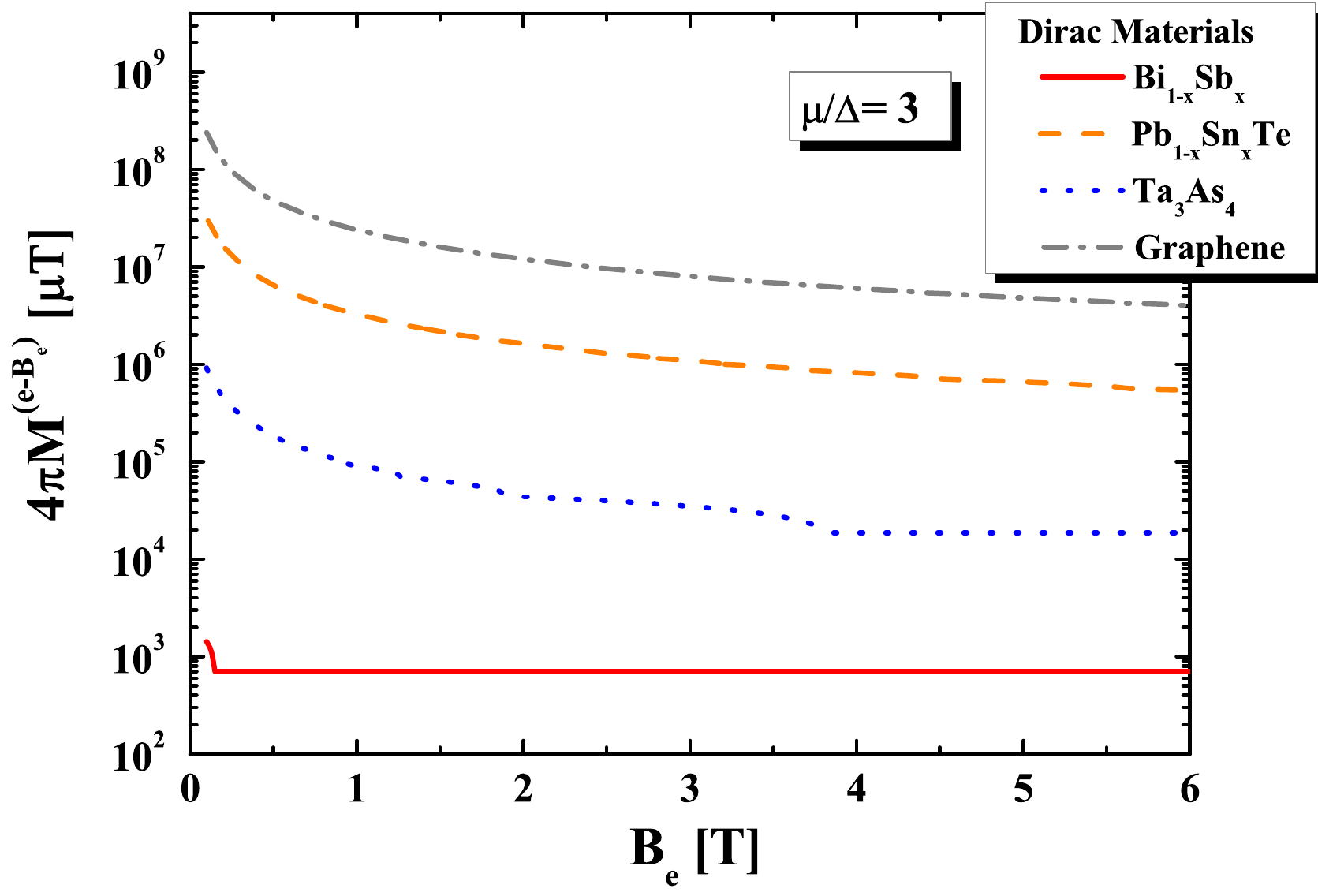}
  \includegraphics[width=0.49\linewidth]{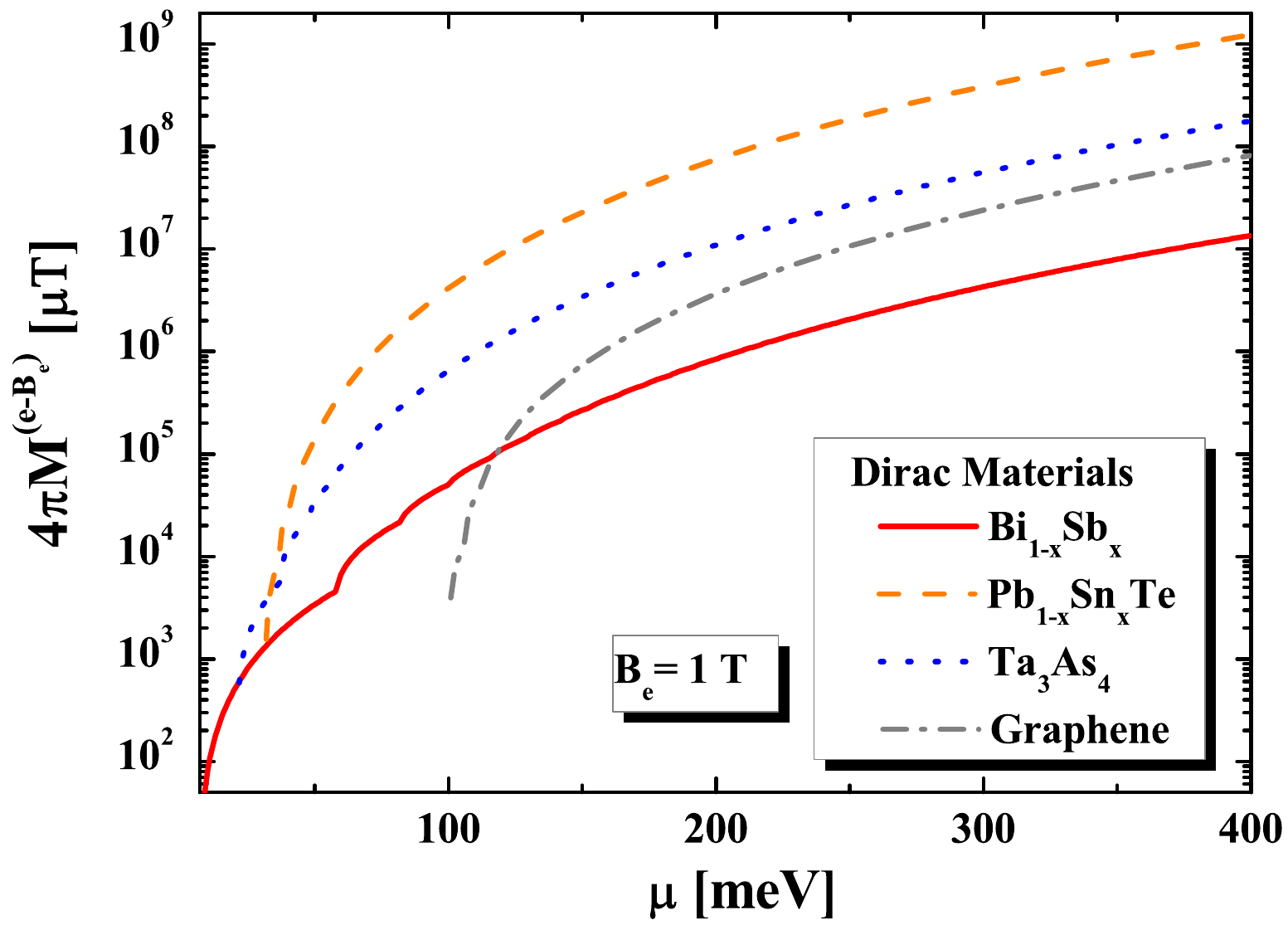}
 \caption{(Left)   Magnetization of electron interacting with magnetic field $\mathcal{M}^{(e-B_e)}$ as a function of the magnetic field $B_e$ in Tesla for a fixed value of $\mu/\Delta=3$. (Right) Electron-field magnetization versus the chemical potential $\mu$ in meV for a fixed value of the external magnetic field $B_e =1 $ T. In both figures have been depicted the magnetization of studied Dirac materials: $Bi_{1-x}Sb_{x} $, $Pb_{1-x}Sn_{x}Te$, $Ta_3 As_4$, and graphene.
 }
\label{mag_med_Tesla}
\end{figure}

Fig. (\ref{mag_med_Tesla}) illustrates the behavior of electron magnetization for the Dirac materials listed in Table I. We have plotted it as a function of magnetic field and chemical potential, revealing similar paramagnetic behavior across all Dirac materials.
In both figures, the oscillations (a characteristic behavior of magnetized quantum gases) arise from the Hass-van Alphen effect, associated with fermion transitions between Landau levels.

The magnetization increases with the band gap and tends to stabilize when the magnetic field exceeds $4 $ T.
Furthermore, magnetization shows a strong dependence on the increase of chemical potential across all materials.
In our model, the magnetization of each material depends solely on the band gap and Fermi velocity. However, it is known that magnetization should also be influenced by the band structure (geometrical factors) and chemical composition. Then, our model tends to overestimate the magnetization, yielding large values $\mathcal{M}^{(e-B_e)}> $ mT. Nevertheless, this value could be a reference for comparison with photon magnetization. 
\begin{figure}[h!]
\centering
\includegraphics[width=0.49\linewidth]{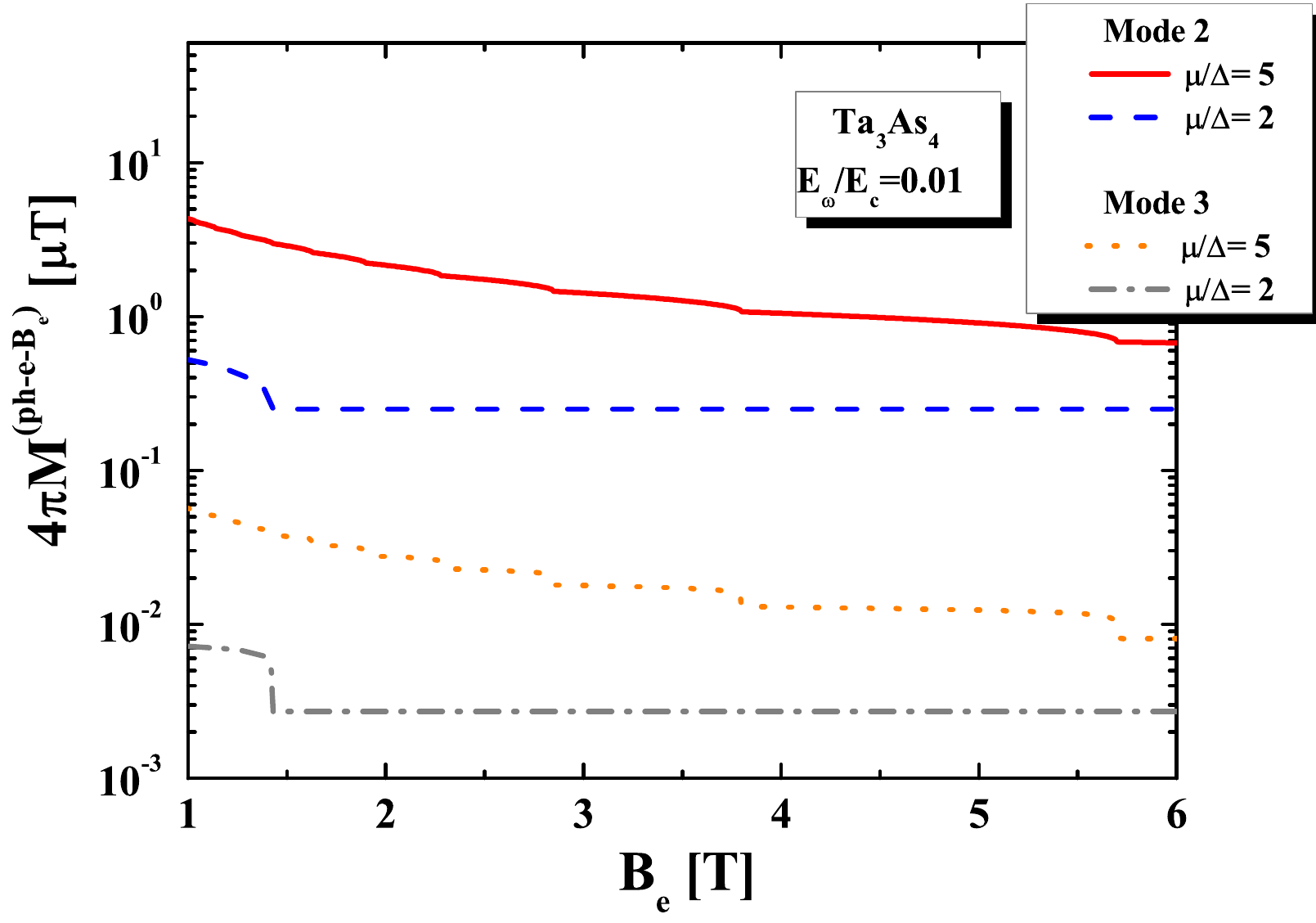}
\includegraphics[width=0.49\linewidth]
{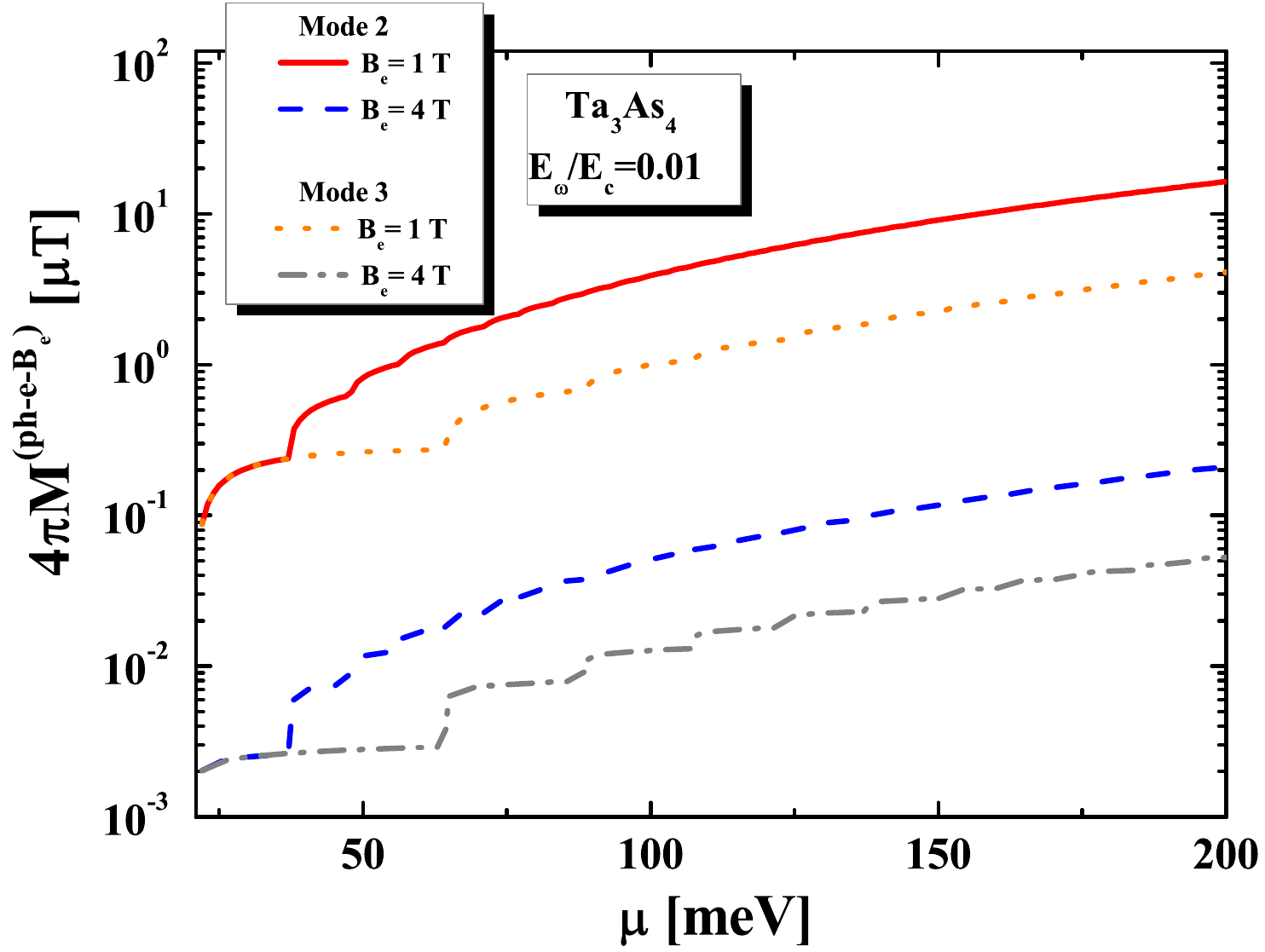}
 \caption{ 
 (Left) Magnetization of photon due to the photon interaction with magnetic field via a medium (electrons) $\mathcal{M}^{(ph-B_e-e)}$ as a function of the magnetic field $B_e$ in tesla for a fixed value of $E_w/E_c=0.01$ with different modes $i=2,3$ and chemical potentials.
 (Right)  Magnetization of the photon interacting with the magnetic field via a medium (electrons) $\mathcal{M}^{(ph-B_e-e)}$ as a function of the chemical potential   $\mu$ in meV for a fixed value of $E_w/E_c=0.01$ with different modes $i=2,3$ and external magnetic fields.
 All figures have depicted the magnetization only for $Ta_3 As_4$.
 }
\label{Magmedioquimicoelectrico}
\end{figure}

The Fig. (\ref{Magmedioquimicoelectrico}) depicts the magnetization of photons interacting with the magnetic field through electrons $\mathcal{M}^{(ph-B_e-e)}$, using Eq. (\ref{mag232}), as a function of the magnetic field and chemical potential specifically for $Ta_3 As_4$. Both figures exhibit the Hass-van Alphen effect, resulting from the quantization of Landau levels.
Similar to the $\mathcal{M}^{(e-B_e)}$, the photon magnetization also increases with the chemical potential. %\textcolor{red}{the heightened density of available charge carriers, such as electrons. With more charge carriers present, there is a greater probability of interaction between these carriers and the magnetic field, resulting in enhanced alignment of magnetic moments and consequently higher magnetization.
%Moreover, fluctuations in the chemical potential can influence the electronic band structure of the material, thereby altering the mobility and behavior of charge carriers, and consequently affecting the material's magnetic properties.}
%
%\textcolor{red}{The photon magnetization in the material constitutes an small  contribution to the material magnetization,  $\mathcal{M}^{(ph-e-B_e)}\ll \mathcal{M}^{(e-B_e)}$. 
Nevertheless, in the figure, we can see that the photon medium magnetization $\mathcal{M}^{(ph-B_e-e)}$ match within the expected range for Dirac materials, from $0.1$ $\mu$T to $10$ $\mu$T \cite{Keser2021yxp, Neves2023uxi}. However, it generally varies depending on factors like the material's Fermi velocity and band gap.

\begin{figure}[h!]
\centering
\includegraphics[width=0.7\linewidth]{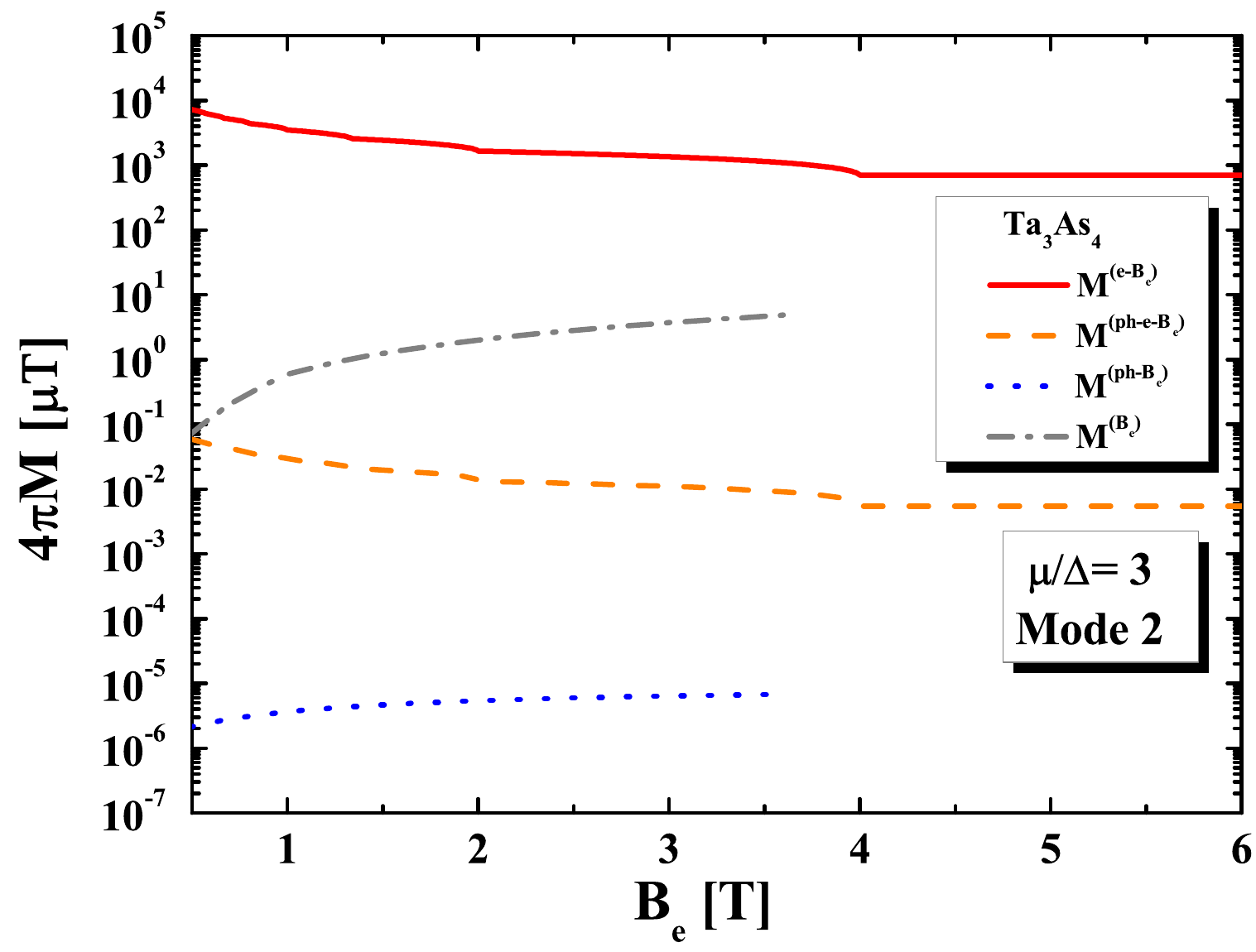}
 \caption{ Comparison of the second polarization mode (2)  magnetization for electron and photon considering both vacuum and medium contribution:  ($\mathcal{M}^{(B_e)}$,   ($\mathcal{M}^{(e-B_e)}$,  $\mathcal{M}^{(ph-B_e)},$  $\mathcal{M}^{(ph-e-B_e)}$) as a function of the external magnetic field in Tesla for a fixed value of chemical potential $\mu/\Delta=3$.  The figures have been depicted for $Ta_3 As_4$.  The value of $E_w$ is normalized using the electrical critical field, $E_\omega/E_c=0.01$, and $E_\omega = v_F B_\omega$.  
 }
\label{MagTOdos}
\end{figure}

Finally, let's compare the photon and electron contributions to magnetization in Fig. (\ref{MagTOdos}). Here, we plot the four contributions of the total magnetization as functions of the external magnetic field for $Ta_3 As_4$.

We observe that the leading contribution of magnetization stems from the electron-field $\mathcal{M}^{(e-B_e)}$ being the most important in the resulting total magnetization. 
%The graphic shows 
The relation between $\frac{\mathcal{M}^{(B_e)}}{\mathcal{M}^{(e-B_e)}}\sim 10^{-3}$ and $\frac{\mathcal{M}^{(ph-B_e)}}{\mathcal{M}^{(B_e-ph-e)}}\sim 10^{-3}$,
%Depending on the parameters of the Dirac material, this magnetization can be three or more orders of magnitude higher than the corresponding photon or vacuum contributions (Fig. \ref{Mag_Vacuum}).
%
%
while the magnetization due to photon propagation is just a tiny quantity in comparison with the material magnetization $\frac{\mathcal{M}^{(ph-B_e)}}{\mathcal{M}^{(e-B_e)}}\sim 10^{-9}$.

%We remark that the magnetization due to photon propagation is just tiny corrections because the photon field is very small compared to the external fields $\frac{\mathcal{M}^{(ph-B_e)}}{\mathcal{M}^{(e-B_e)}}\sim 10^{-9}$. 
To measure this quantity, one requires greater precision and new techniques, such as those provided by superconducting quantum interference device (SQUID) magnetometry, whose sensitivity can reach up to $10^{-15}$ T/Hz$^{1/2}$ \cite{Keser2021yxp, SQUID}. We hope that these techniques will aid in the detection of vacuum magnetization. However, separating vacuum and medium contributions remains a challenge.

\section{Effective photon magnetic moment}

Let us study the effective magnetic moment of a photon probe propagating in a magnetized Dirac vacuum, defined from the variation of the magnetic energy $\mathcal{E}=\mathcal{M}^{(ph-B_e)} B_e=\mathcal{L}_{\mathcal{FF}}\langle{E}_w^2\rangle B_e^2 +\mathcal{L}_{\mathcal{GG}}\langle{E}_w^2\rangle B_e^2$
of the photon propagating in a Dirac vacuum with respect to the magnetic field. We translate to Dirac material language the results obtained in \cite{PerezGarcia2022kvz} 
%\begin{equation}
%\mid \pmb{\mu}_{ph}\mid=-\frac{d \langle \mathcal{E^D}_{mag} %\rangle_V } {d B_e} \frac{1}{V\langle N_V\rangle},
%\end{equation}
%where $N_V^{(i)}$ is the number density of ith mode  ($i=2,3$) and is given by  $N_V^{(2,3)}= \frac{1}{2}\frac{E_0^2}{\omega^{(2,3)}}$ \cite{Villalba-Chavez:2012pmx}. $\pmb{\mu}_{ph}$ will characterize the interaction of propagating photons with the vacuum virtual pairs under the presence of a magnetic field.
The explicit expression for the two modes $\mid \pmb{\mu}_{ph}^{(2,3)} \mid$ reads as

\begin{align}
&\mid\pmb{\mu}_{ph}^{(2)}\mid =\dfrac{\alpha_D}{16\pi}\dfrac{1}{b^3} \left\lbrace
3-12 \zeta^{(1,1)}\left( -1,\dfrac{1}{2b} \right )+3\psi\left( \dfrac{1}{2b} \right )\right\rbrace\nonumber\\
&+ b \left[-3 +\log  \Gamma \left(\dfrac{1}{2b}  \right ) \left (\dfrac{\pi}{b}
\right )^2\right. +\left.\left.\psi^{(1)}\left (1+\dfrac{1}{2b}\right )+ 2b^2\right ] \right\rbrace\dfrac{\mid\bf{k}_{\bot}\mid}{B_c},
\end{align}

\begin{align}
&\mid \pmb{\mu}_{ph}^{(3)}\mid =\dfrac{\alpha_D}{8\pi}\dfrac{1}{b^4}\left\lbrace
	-\psi^{(1)}\left (1+\dfrac{1}{2b}\right ) \right.
	+ b \left [	4-	4\psi\left (1+\dfrac{1}{2b}\right )+2\psi\left (\dfrac{1}{2b}\right )
	\right  ] \nonumber\\
&+b^2 \left[4-2 \log(2\pi)
+\left.\left. 4\log\left (\Gamma \left( \dfrac{1}{2b}  \right ) \left (\dfrac{\pi}{b}  \right )^{1/2}\right ) \right] \right\rbrace \dfrac{\mid{\bf{k}_{\bot}}\mid}{B_c},  \right.
\end{align}

\begin{figure}[h]
\centering
\includegraphics[width=0.49\linewidth]{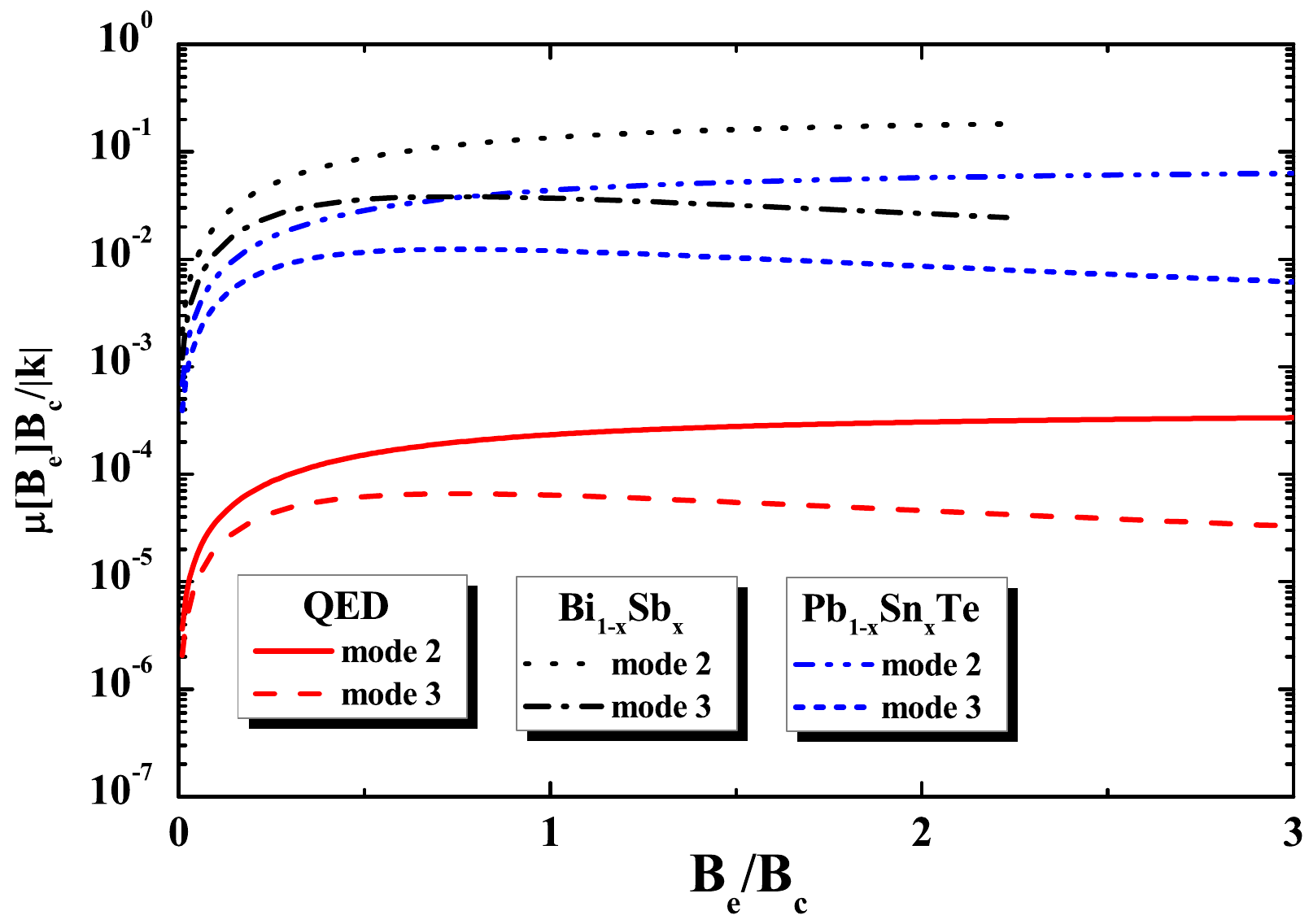}
\includegraphics[width=0.49\linewidth]{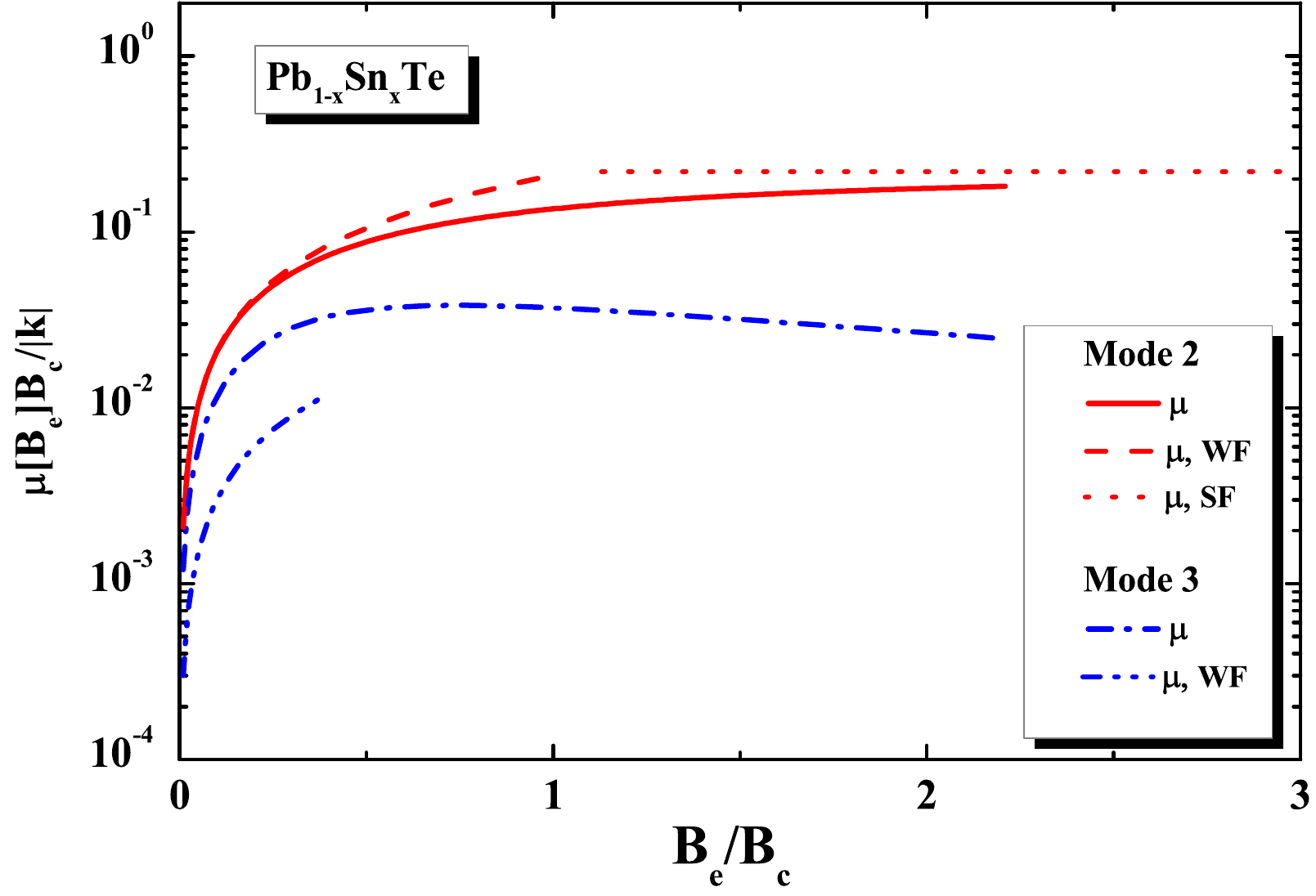}
 \caption{Photon effective magnetic moment as a function of magnetic field strength and $B_e/B_{c}$. The left figure shows the magnetic moment for two Dirac materials: $Pb_{1-x}Sn_{x}Te$ and  $Bi_{1-x}Sb_{x} $. We have plotted the corresponding one for QED for both modes. For each case, the modes (2) and (3) are depicted. The right figure shows for  $Pb_{1-x}Sn_{x}Te$ the comparison of the magnetic moment for an arbitrary value of the magnetic field with the corresponding two limits weak (dashed and dotted red line) and strong (dashed-dot-dot blue line) field limits for both modes respectively. }
\label{fig:MMmoment}
\end{figure}
where $\psi^{(1)}=\partial_h\psi[h]$,  $\psi$ is the PolyGamma o Digamma function, (first derivative of ${\rm ln}\, \Gamma$). $\zeta^{(1,1)}[s,h]=\partial_h \zeta^{\prime}$ with $\zeta^{\prime}=\partial_s\zeta[s,h]$ and $\zeta[s,h]$ is the Hurwitz zeta function  \cite{Adam}.
Considering the form of $\psi$ and  $\zeta^{\prime}$ in weak and strong field limits (see details in \cite{PerezGarcia2022kvz}). For  modes (2) and (3)  yields
$\mid \mu_{ph}^{WF\,(2)}\mid =\frac{\alpha_D}{4\pi}\frac{28}{45}\frac{B_e}{B_c^2}\mid{\bf k}_{\bot}\mid$ and
%\frac{\alpha^2}{4\pi}\frac{28}{45}\frac{k_{\bot}}{m^2}
$\mid \mu_{ph}^{WF\,(3)}\mid= \frac{4}{7} \mid \mu_{ph}^{WF\,(2)}\mid$.
While for strong field limit, only  mode (2) contributes to the effective magnetic moment, and it goes to a constant value, % Furthermore, the anomalous electron magnetic moment, $\mu_e$,
$\mid \mu_{ph}^{SF\,\,(2)}\mid=\frac{\alpha_D}{3\pi}\frac{\langle E_w^2 \rangle}{B_c}$
%&\sim \frac{\alpha_D }{3\pi}\frac{e}{2m}\frac{\mid {\bf k}_{\bot}\mid}{m}$
and $\mid \mu_{ph}^{SF\,\,(3)} \mid=0.$ Note that the effective photon magnetic moment $\mid \mu_{ph}^{SF\,\,(2)} \mid \sim 10^{-1} \mu_e$ is a decimal of the electron magnetic moment ($\mu_e$), contrasting to the QED value,  which is two orders lower.

In Fig. (\ref{fig:MMmoment}) left, we depict the photon's effective magnetic moment as a function of $b$ for polarization modes (2) and (3) of Dirac materials.
For comparison, we have depicted the magnetic moment for light propagating in a magnetized vacuum of QED.
One can see that for the Dirac material, the effective magnetic moment for   $Pb_{1-x}Sn_{x}Te$ is one order below the QED magnetic moment, while for $Bi_{1-x}Sb_{x} $ is three orders lower than the value of QED.
On the right side of Fig. (\ref{fig:MMmoment}), we plot for  $Pb_{1-x}Sn_{x}Te$ the effective magnetic moment, considering the results for arbitrary values of the magnetic field as well as we have plotted it for weak and strong limit for both modes.
%As can be seen, the %effective moment is over %(under) estimated if we take the approximate values arising from the weak and strong limits in their range of validity.
Then, for a magnetic field, $B_{e}\gtrsim 2B_{c}$ the effective magnetic moment of photons polarized on mode (2) tends asymptotically to a constant value \cite{Mielniczuk, Villalba-Chavez:2012pmx, Elizabeth}. Mode (3) slowly decreases with the magnetic field strength, being zero, its strong field limit value.

%%%%%%
\section{Dynamics properties of photon propagating in a Dirac vacuum}

We study in this section the equation of motion  of the photon traveling in the Dirac vacuum $\mathcal{L}_D^{(ph-B_e)}$ using the minimum action principle, it reads as
\begin{equation}
\partial_{\mu} \left(\frac{\partial\mathcal{L}^D}{\partial f_{\mu\nu}} \right) = -\frac{1}{2}(1-\mathcal{L}^D_\mathcal{F})
\partial_{\mu}f^{\mu\nu} +\partial_{\mu}
\left(\frac{\mathcal{L}^D_{\mathcal{FF}}}{2}(F^{\sigma\rho}f_{\sigma\rho}(F^{\mu\nu}))
+\frac{\mathcal{L}^D_{\mathcal{GG}}} {2}(\tilde{F}^{\sigma\rho}f_{\sigma\rho}(\tilde{F}^{\mu\nu}) \right)=0, \label{EqMovlineal2}
\end{equation}

\noindent where the dual tensor $\tilde{f}_{\mu\nu}$  satisfies the Bianchi equation  $\partial_{\mu} \tilde{f}_{\mu\nu}=0$,  or second pair of Maxwell equation  for ${\bf E}_w$ and ${\bf B}_w$,
\begin{align}
\nabla \cdot \mathbf{B}_w=0,\quad \quad
\frac{\partial \mathbf{B}_w}{\partial t}=-\nabla \times  \mathbf{\bf{E}}_w, \label{magpo2}
\end{align}
with $i=1,2,3$.
\noindent and the modified Maxwell equation in terms of constitutive vectors
$\mathbf{D}_w$ and $\mathbf{H}_w$  are
\begin{align}
\frac{\partial\mathbf{D}_w}{\partial t}=-\nabla \times \mathbf{H}_w,   \quad \quad \nabla \cdot \mathbf{D}_w=0, \label{magpo1}
\end{align}
where
\begin{align}
D_{w,i}=
%\frac{\partial}{\partial E_{w,i}}[ \mathcal{L}^{\rm (ph-B_e)}]=
\epsilon_{ij}(B_e)E_{w,j}, \quad  \quad
H_{w,i}=-%\frac{\partial}%{\partial B_{w,i}}[ %\mathcal{L}^{\rm (ph-B_e)}]=
(\mu^{-1})_{ij}(B_e){B}_{w,j}.
\end{align}

\begin{figure}[h]
\centering    \includegraphics[width=0.7\linewidth]{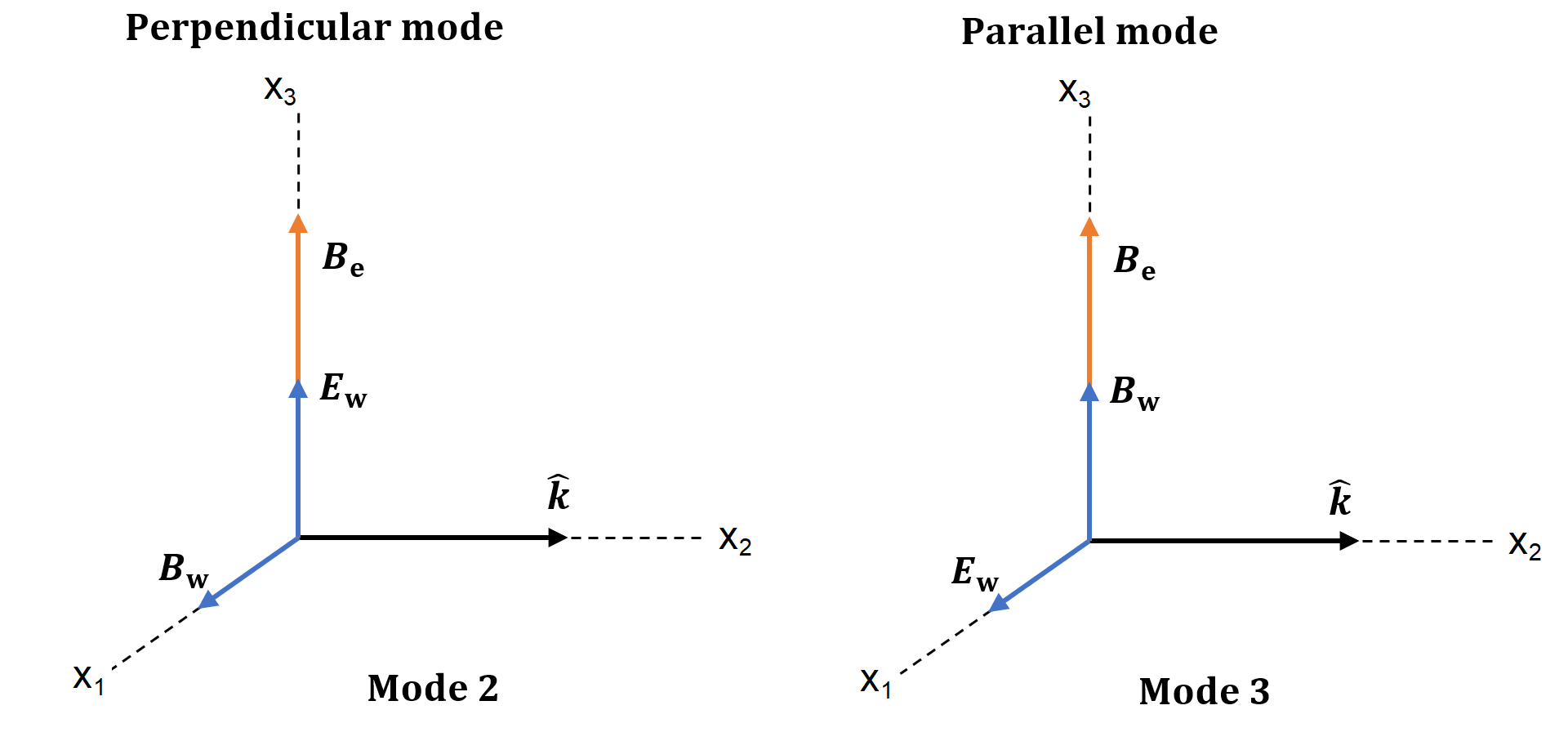}
 \caption{ Two physical transverse polarization modes of the photon field, mode (2), $E_w\parallel B_e$ and mode (3), $B_w\parallel B_e$. The external magnetic field $B_e$ in  $\pmb{\hat{x_3}}$ direction $\textbf{B}_e=B_e\pmb{\hat{x_3}}$,  and the photon propagates in $\pmb{\hat{x_2}}$.
 }
\label{fig:modes}
\end{figure}

In our study, we are considering an external magnetic field in  $\pmb{\hat{x_3}}$ direction $\textbf{B}_e=B_e\pmb{\hat{x_3}}$,  and the photon propagates in $\pmb{\hat{x_2}}$ (Fig. \ref{fig:modes}). In that case, only non-zero components of the electric permittivity and magnetic permeability tensors, are  

\begin{equation}
\epsilon_{\bot}=\mu_{\bot}=1-\mathcal {L}_{\mathcal F},\label{epsilon}\quad \quad
\epsilon_{\parallel}=(1-\mathcal {L}_{\mathcal F}+2{\mathcal F}\mathcal {L}_{\mathcal {GG}}),
\end{equation}
\begin{equation}
\mu_{\parallel}=(1-\mathcal {L}_{\mathcal F}-2{\mathcal F}\mathcal {L}_{\mathcal {FF}}).\label{epsimu2}
\end{equation}
\subsection{Dispersion equation of photon propagating in a Dirac vacuum}

Considering the photon propagation transverse to the magnetic field (Fig. \ref{fig:modes})  of a plane wave  as ${\bf E}_w={\bf E}_0 \rm e^{-i(\mathbf{k_{\perp}}\cdot\mathbf{x_2}-\omega t)}$, from the Maxwell equations we can obtain the dispersion equation of photon as
\begin{equation}
(k_{\bot}^2\mu_{\bot\,\parallel}^{-1}+\omega^{2}\epsilon_{\parallel\,\bot})E^{\parallel\,\bot}_{w}=0.\label{ecdis}
\end{equation}

\begin{figure}[h!]
\centering    \includegraphics[width=0.65\linewidth]{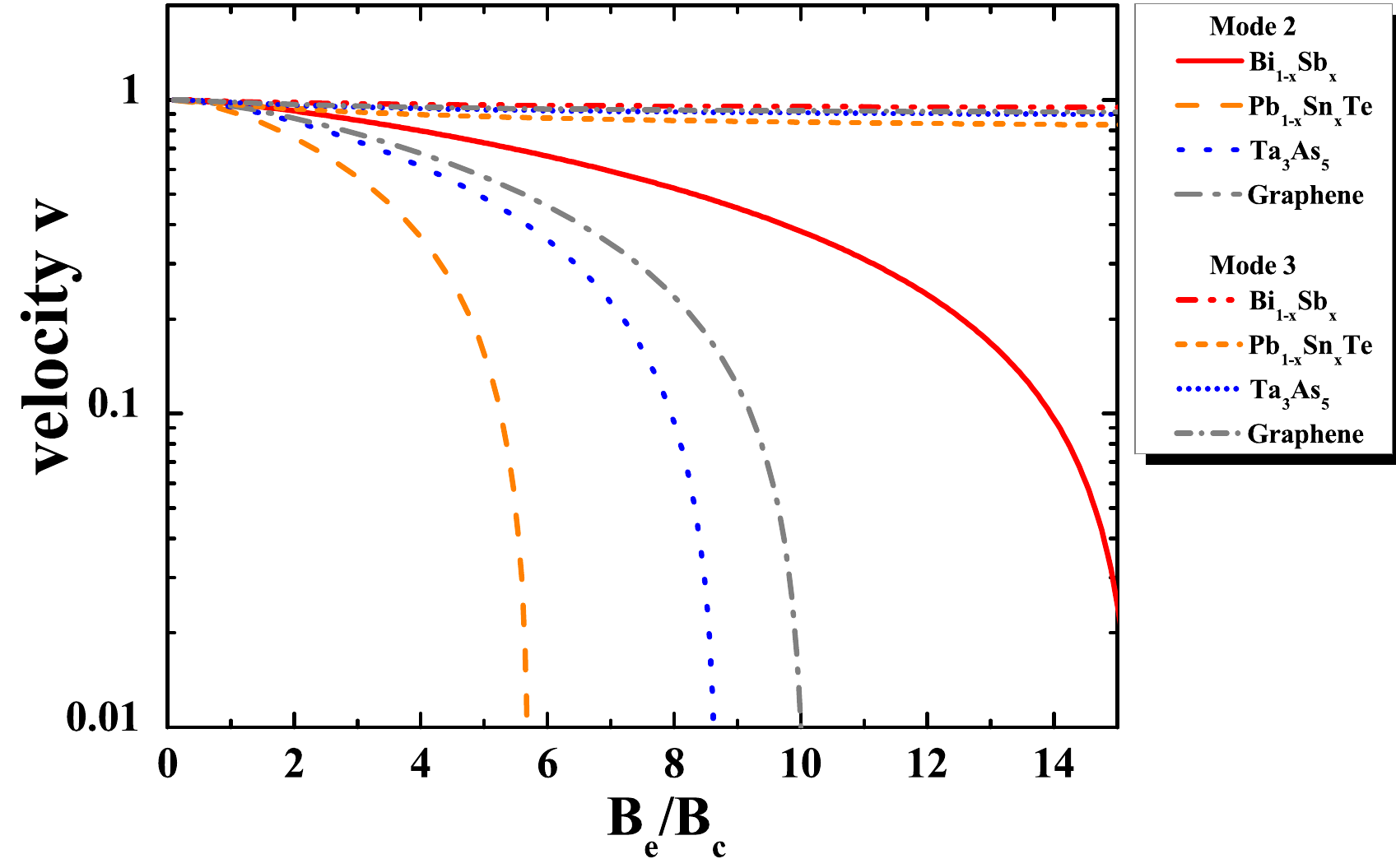}
 \caption{Phase velocity for both second and third mode of propagation of the photon as a function of the dimensionless magnetic field. The non-linear effect decreases the phase velocity with the $B_e$ increase. We have plotted curves for four Dirac materials:  $Bi_{1-x}Sb_{x} $, $Pb_{1-x}Sn_{x}Te$, $Ta_3 As_4$, and graphene. As we can see, the lowest velocity is obtained for  $Pb_{1-x}Sn_{x}Te$ (dashed and short dashed orange) due to its effective fine structure constant being the highest.}\label{fig:velocityphase}
\end{figure}
%\begin{equation}
%(\epsilon_{ijk}\epsilon_{lab}k_{j}(\mu^{-1})_{kl}k_a+\omega^{2}\epsilon_{ib})E_{w\,b}=0,\label{ecdis}
%\end{equation}

%\noindent where $i,j,k, a,b,l=1,2,3$.

The solution of Eq. (\ref{ecdis}) describes two physical transverse polarization modes of the photon field, mode (2) where the $E_w\parallel B_e$ and mode (3) $B_w\parallel B_e$.

The dispersion relation for each polarization mode has the form
\begin{align}
\omega^{(2)} \simeq \mid {\bf k}_{\bot} \mid \left(1-\frac{{\mathcal L}^D_{\mathcal {GG}}B_e^2}{2}\right),  \quad \quad
\omega^{(3)} \simeq \mid{\bf k}_{\bot} \mid \left(1-\frac{{\mathcal L}^D_{\mathcal {FF}}B_e^2}{2}\right),
\end{align}
\noindent in agreement  with \cite{Elizabeth,Shabad:2011hf,romero2020photon}. The appearance of Cotton-Mouton birefringence \cite{Rizzo:2010di} brings the existence of two refraction indexes associated with the two different polarization modes: $n_{\parallel}$ for mode (2)  and $n_{\bot}$ for mode (3),

\begin{equation}\label{indrefraccion}
n_{\parallel,\bot}=\frac{\mid \bf{k_{\bot}}\mid }{\omega^{(2,3)}}=\sqrt{\frac{\epsilon_{\parallel,\bot}}{\mu_{\bot,\parallel}}}.
\end{equation}

\noindent The difference between the refraction index $\Delta n= n_{\parallel}-n_{\bot}$ takes the form
\begin{equation}
\Delta n=\frac{(\mathcal{L}^D_{\mathcal{GG}}-\mathcal{L}^D_{\mathcal{FF}})B_e^2}{2}.\label{birefringence}
\end{equation}
 In the weak field limit,  it is reduced to  $\Delta n^{\rm WF}_{CM}=3/4\xi_D B_e^2$, instead of the strong limit $\Delta n^{\rm SF}_{\rm CM}=\frac{\alpha_D}{3\pi}(\frac{B_e}{B_{c}}-1)$. Since the phase velocity
 $v_{\parallel,\bot}=1/n_{\parallel,\bot}$, in the strong field limit the condition, $v_{\bot}^{SF}=1-\frac{\alpha_D}{3\pi}(\frac{B_e}{B_{c}})>0$   fixes the validity of one-loop approximation up to values of the magnetic field  \cite{DittrichLibro} $b\leq 3 \pi/\alpha_D $.

The phase velocity as a function of the magnetic field is plotted for a second mode for four Dirac materials in Fig. (\ref{fig:velocityphase}).    We have depicted curves for $Bi_{1-x}Sb_{x} $, graphene, $Ta_3 As_4$, and $Pb_{1-x}Sn_{x}Te$. The graphic shows that $Pb_{1-x}Sn_{x}Te$ has the lowest velocity with the high effective fine structure constant.
The dispersion equation in weak  magnetic field limit \cite{Elizabeth, Hugo1,PerezGarcia2022kvz,perez2023photon} is obtained as
\begin{align}
\omega^{WF,(2)}\simeq\mid\boldsymbol{k_\perp}\mid \left(1-\frac{7}{4}\xi_D B_e^2 \right),\quad
\omega^{WF,(3)}\simeq\mid\boldsymbol{k_\perp}\mid (1-\xi_D B_e^2),\label{Bperk23}
\end{align}
\noindent while for strong magnetic field limit, the result is
\begin{align}\label{omegaS}
\omega^{SF,(2)}
%=\mid\boldsymbol{k}\mid
%(1-\frac{\alpha}{3\pi}
%(\boldsymbol{\hat{b}\times\hat{k}}))
\simeq \mid\boldsymbol{k_\perp}\mid \left (1-\frac{\alpha_D}{3\pi}\frac{B_e}{B_{c}}\right ),\quad
\omega^{SF,\,(3)}\simeq\mid \boldsymbol{k_\perp}\mid
%(1-\frac{\alpha}{3\pi}\frac{B}{B_{c}} (\boldsymbol{\hat{b}\times\hat{k}}))
%=\mid\boldsymbol{k}\mid
\left (1-\frac{\alpha_D}{3\pi}\right).
\end{align}
\section{Photon Energy-momentum tensor: energy density, Pointing vector and radiation pressures}

The energy-momentum tensor (EMT)  was calculated and widely discussed in our previous work \cite{PerezGarcia2022kvz} for photon propagating in a magnetic field. It was obtained from the effective Euler Heisenberg Lagrangian of QED  by Hilbert method which by construction provides a symmetric EMT tensor. Besides, was proved that the EMT is gauge invariant and conserved \cite{PerezGarcia2022kvz}. 
Hilbert's method for calculating the EM tensor consists of varying the effective Lagrangian 
% $\mathcal{L}^{ph-B_e}_D$
with respect to the metric tensor $g^{\mu\nu}$ and then recovering the flat space doing  
$g^{\mu\nu}\rightarrow\eta^{\mu\nu}$ Euclidean metric.   
%In \cite{PerezGarcia2022kvz}, we have also clarified why some authors obtained a non-symmetric EM tensor when using the Noether-Belfinante method  \cite{Villalba-Chavez:2012pmx}.   
%The problem lies in that when calculating the Energy-Momentum Tensor (EMT) using the Noether-Belifante method, it's necessary to derive the Lagrangian $\mathcal{L}^{ph-B_e}$, which has already been approximated by a series expansion. As a result, some terms are missing when compared to the tensor found by the Hilbert method. 
%However an equivalent result is obtained if the starting point to get EMT for Noether method is a general Lagrangian $\mathcal{L}=\mathcal{L}(\mathcal{F},\mathcal{G})$ with $\mathcal{F}_{\mu\nu}={F}_{\mu\nu}+{f}_{\mu\nu}$.  Once EMT is computed it should be expanded up to second order in the photon field ${f}_{\mu\nu}\ll {F}_{\mu\nu}$ yielding an equivalent EMT that for Hilbert method. Details of all the discussion and calculations are given in \cite{PerezGarcia2022kvz}}
% $g^{\mu\nu}$ in the usual way. 
%It has two main %contributions
%
%\begin{equation}
%T_H^{\gamma\rho} %=T_H^{(0) \gamma\rho} + t_H^{\gamma\rho},
%T_H^{(1) \gamma\rho} +T_H^{(2) \gamma\rho},
%\end{equation}
%where $T_H^{(0) %\gamma\rho}$ is that %of the external %background field 
Proceeding in a similar way that in \cite{PerezGarcia2022kvz} we get the EMT   from the photon Lagrangian $\mathcal{L}_D^{(ph-B_e)}$.
$t_{H}^{\gamma\rho}$ accounts for the interaction of the photon field with the background field in a Dirac vacuum, it has the form

\begin{equation}
t_{H}^{\gamma\rho} =\left.\frac{2}{\sqrt{-g}} \frac{\delta \mathcal{L}_D^{(ph-B_e)}}{\delta g_{\gamma \rho}}\right|_{g=\eta}.
\label{metricder}
\end{equation}

Performing the derivatives and recovering the flat space, 
%and extrapolating 
its form to Dirac material yields
\begin{align}
t^{H\,\,\gamma\rho}&=(1-\mathcal{L}^D_{\mathcal{F}})f^{\gamma}_{\lambda}f^{\lambda\rho}+\frac{\mathcal{L}^D_{\mathcal{FF}}}{2}f^{\mu\nu}F_{\mu\nu}(F^{\gamma\alpha}
f_{\alpha}^{\rho}+F^{\rho\alpha}f_{\alpha}^{\gamma})
\nonumber\\
&+\frac{\mathcal{L}^D_{\mathcal{GG}}} {2} f^{\mu\nu} \tilde{F}_{\mu\nu}(\tilde{F}^{\gamma\alpha}f_{\alpha}^{\rho}
+\tilde{F}^{\rho\alpha}f_{\alpha}^{\gamma})+\frac{\eta^{\gamma\rho}}{4}\left((1-\mathcal{L}^D_{\mathcal{F}})
f_{\mu\nu}f^{\mu\nu}\right.\nonumber\\
&+\frac{\mathcal{L}^D_{\mathcal{FF}}}{2}f^{\mu\nu}F_{\mu\nu}f^{\alpha\beta}F_{\alpha\beta}+\frac{\mathcal{L}^D_{\mathcal{GG}}} {2}\left. f^{\mu\nu} \tilde{F}_{\mu\nu}f^{\alpha\beta}\tilde{F}_{\alpha\beta}\right)\nonumber\\
&+\mathcal{L}^D_{\mathcal{F}} (F^{\gamma\alpha}f_{\alpha}^{\rho}+F^{\rho\alpha}f_{\alpha}^{\gamma})+\frac{\eta^{\gamma\rho}}{2}\mathcal{L}^D_{\mathcal{F}}f^{\mu\nu}F_{\mu\nu}.
\label{defEMT1}
\end{align}

The external magnetic field makes EMT fully anisotropic \cite{PerezGarcia2022kvz}. The average of the diagonal part of the tensor corresponds to energy density and anisotropic pressures, while the non-diagonal components account for the Poynting vector.

%Let us note that Hilbert method provides an EMT

%$$t^{H\,\,\gamma\rho}=t^{N\,\,\gamma\rho} +\frac{\mathcal{L}_{\mathcal{FF}}}{2}\mathcal{F}^{\rho\alpha}f_{\alpha}^{\gamma}
%\tilde{F}^{\rho\alpha}f_{\alpha}^{\gamma} +\frac{\mathcal{L}_{\mathcal{GG}}} {2} f^{\mu\nu} \tilde{F}_{\mu\nu}
%\tilde{F}^{\rho\alpha}f_{\alpha}^{\gamma},$$

The energy density $\langle t^{00}\rangle$ for both modes is positive \cite{Neves2023uxi}, and  the explicit expression for photon energies for both modes are,
\begin{align}
\mathcal{E^D}_{w}^{(2)}&\simeq (1-\mathcal{L}^D_{\mathcal{F}}+{\frac{3}{2}}\mathcal{L}^D_{\mathcal{GG}}B_e^2)\langle E_{w}^2\rangle,\quad \quad
\mathcal{E^D}_w^{(3)}\simeq (1 -\mathcal{L}^D_{\mathcal{F}}-{\frac{1}{2}}\mathcal{L}^D_{\mathcal{FF}}B_e^2)\langle E^{2}_{w}\rangle.
\end{align}
However, as the Hamiltonian of the effective theory is
\begin{align}
\mathcal{H^ D}=D_wE_w- \mathcal{L}_D^{\rm (ph-B_e)} =\frac{(1-\mathcal{L}^D_{\mathcal{F}})}{2}(E_w^2-B_w^2)-
\frac{\mathcal{L}^D_{\mathcal{FF}}}{2}({\bf B} \cdot{\bf B}_w)^2+\frac{\mathcal{L}^D_{\mathcal{GG}}}{2}({\bf B}\cdot {\bf E}_w)^2,
\end{align}
only the energy density for mode (3) becomes in the eigenvalue of the Hamiltonian \cite{Neves2023uxi}.

The anisotropic pressures take the form
\begin{align}
p^{D\,\,(2)}_{1,w}&=(1-\frac{1}{2}\mathcal{L}^D_{\mathcal{GG}}B_e^2)\langle E_{w}^2\rangle,\quad\quad
p^{D\,\,(3)}_{1,w}=(1-\frac{1}{2}\mathcal{L}^D_{\mathcal{FF}}B_e^2)\langle E_{w}^2\rangle,\nonumber \\
p^{D\,\,(2)}_{2,w}&=(1-\mathcal{L}^D_{\mathcal{F}}+\frac{\mathcal{L}^D_{\mathcal{GG}}B^2}{2})\langle E_{w}^2\rangle,\quad\quad
 p^{D\,\,(3)}_{2,w}=(1-\mathcal{L}_{\mathcal{F}}-\frac{3}{2}\mathcal{L}^D_{\mathcal{FF}}B^2)\langle E_{w}^2\rangle,\label{pressurey} \\
 p^{D\,\,(2)}_{w,\parallel}&\simeq(1-\frac{3}{2}\mathcal{L}^D_{\mathcal{GG}} B_e^2)\langle E_w^2\rangle,  \quad\quad
p^{D\,\,(3)}_{w,\parallel}\simeq(1+\frac{1}{2}\mathcal{L}^D_{\mathcal{FF}}B_e^2)\langle E_{w}^2\rangle, \label{presure2}
\end{align}
while the Poynting  vector  has the form,
\begin{align}
\mathcal{P^D}_w^{(2)}&\simeq(1-\mathcal{L}^D_{\mathcal{F}}+\mathcal{L}^D_{\mathcal{GG}}B_e^2)\langle E_{w}^2\rangle,\quad\quad\label{Poynting}
\mathcal{P^D}_w^{(3)}
\simeq(1-\mathcal{L}^D_{\mathcal{F}}-\mathcal{L}^D_{\mathcal{FF}}B_e^2)\langle E_{w}^2\rangle.\nonumber\\
\end{align}
We can rewrite the above quantities in terms of vacuum photon magnetization. In particular, the Poynting vector takes the form
\begin{equation}
\mathcal{P^D}_w^{(2,3)}=P_0\pm \frac{1}{2}\mathcal{M}_D^{ph-B_e(2,3)}B_e,\label{Poynting}
\end{equation}
where $P_0$ contains the non-linear correction that emerges from  the scalar invariant $\mathcal{F}$ with, $P_0=(1-\mathcal{L}^D_{\mathcal{F}})
\left\langle E_w^2\right\rangle$ while the second term comes from the pseudo-scalar invariant $\mathcal{G}$ and it is proportional to the magnetization (\ref{mag23}). 
%and

%\begin{equation}
%\mathcal{M}_w^{(2)}=\mathcal{L}_{\mathcal{GG}}B_e \left\langle E_w^2\right\rangle,  \quad \quad \mathcal{M}_w^{(3)}=\mathcal{L}_{\mathcal{FF}}B_e \left\langle E_w^2\right\rangle.  \label{mag23}
%\end{equation}

%The radiation pressure could be defined by the Poynting vector(\ref {Poynting1}) and/or as the target pressure $p_{w}^2$ Eq.(\ref{presure2})  ($\pmb{{\hat y}}$-direction component) as
%

\begin{figure}[h!]
\centering
  \includegraphics[width=0.58\linewidth]{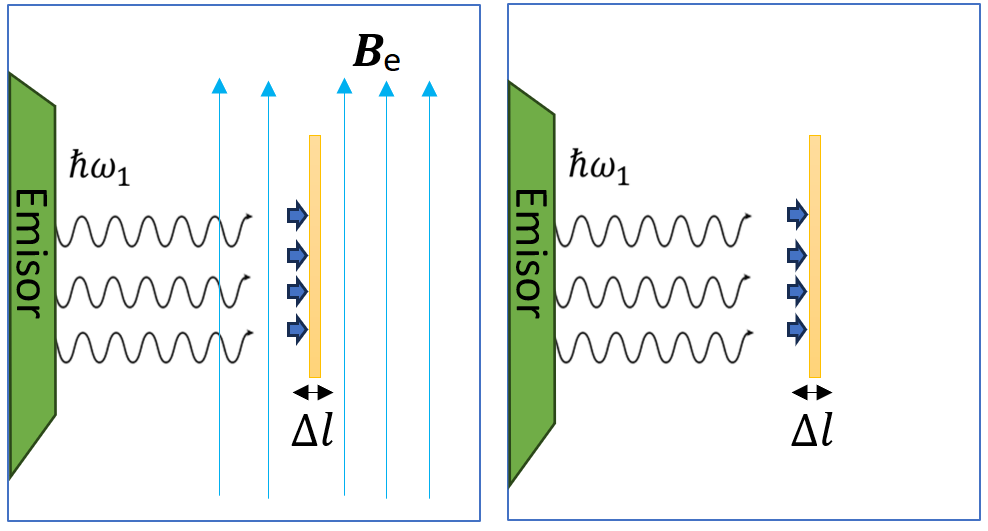}
 \caption{Proposed experiment for measuring the pressure of the radiation over the Dirac material with (left) and without (right) external magnetic field. The radiation pressures over the sheet could be measured using Eq. (\ref{Poynting}). }
\label{Exp_Presiones}
\end{figure}

Let us highlight that Classical Electrodynamics as well as non-linear isotropic electrodynamics with Lagrangian only dependently on the scalar invariant  $\mathcal{F}$, lead to the target pressure $p_w^2$ to be equal to the Poynting vector.  
However, the presence of the magnetic field in EH non-linear Lagrangian involves both invariant  $\mathcal{F}$ and the pseudo-scalar invariant $\mathcal{G}$. As we have discussed before, $\mathcal{G}$ leads to anisotropies due to the rotation symmetry being broken. Besides, it also leads to $p_w^2\neq\mathcal{P}_w$.  %However, the Pointyng theorem $\frac{\partial p_w^2}{\partial x_2} =-\frac{\partial {\mathcal{P}_w}}{\partial t} $.% fullfils the Pointyng vector and the pressure is that  $\partial_{2} p_{2}=\partial_{2} \mathcal{P}_w$ .

On the other hand, the radiation pressure behaves differently depending on their propagation mode. For mode (2) it is higher than the corresponding “classical pressures” while for mode (3), it becomes lower than the classical value. 
%
%\textcolor{red}{At first glance, the behavior of the radiation pressure for mode (2) seems contrary to intuition because the phase velocity is lower than if it propagates in an “empty vacuum” $v_F=1-\frac{\mathcal{L}^D_{\mathcal{GG}}}{2}B_e^2$. The contribution of magnetic energy  enters the radiation pressure as an additive term $\mathcal{P^D}_w^{(2)}\simeq(1-\mathcal{L}^D_{\mathcal{F}}+\mathcal{L}^D_{\mathcal{GG}}B_e^2)\langle E_{w}^2\rangle$, increasing the pressure with respect to the classical one, see Eq. (\ref{presure2}-\ref{Poynting}). Instead, for mode (3), this contribution appears subtracting, yielding thus a lower pressure value, see Eq. (\ref{presure2}).}

%\textbf{A cursory observation might suggest that the behavior of mode (2) is counterintuitive. Despite a photon traveling through a magnetized vacuum exhibiting a slower velocity compared to one in an ``empty vacuum," the magnetic energy's contribution enhances the radiation pressure as an additional term. This augmentation contrasts with classical expectations, as depicted in Eq. (\ref{presure2}-\ref{Poynting}). Conversely, for mode (3), this contribution appears as a deduction, resulting in a reduced pressure value, as shown in Eq.(\ref{presure2}).}

%\subsection{Experiments for vacuum in Dirac materials}

Let's imagine an experiment to detect non-linear radiation pressure. For this purpose, let's assume a monochromatic light emitter (laser), a polarizer that allows polarization in both physical transverse modes, an external magnetic field $\mathbf{B}_e$ in the direction $\hat{x_3}$ of the order of $B_e=0.5B_c$, and a sheet of $Ta_3 As_4$ as a 3D Dirac material of thickness $\Delta l$. The sheet is located parallel to the magnetic field in the  $x_1 x_3$ plane. The pressure of the incident beam on the sheet, %depending on the photon polarization,
would depend on $\mathcal{L}^D_{\mathcal{GG}}$ or $\mathcal{L}^D_{\mathcal{FF}}$ for a chosen polarization beam being greater or lower than the radiation pressure exerted by the same beam when there is no magnetic field, Fig. (\ref{Exp_Presiones}). The non-linear pressure correction will be proportional to the magnetization. Considering  the laser beam intensity, $I\sim\langle E_w^2\rangle $ the radiation pressure will be $|\mathcal{P^D}_w^{(2)}/I-1|\approx 2.1 \times 10^{-2}$ for mode (2) and $|\mathcal{P^D}_w^{(3)}/I-1| \approx 2.2 \times 10^{-2}$ for mode (3). In the context of QED, it's notable that the correction factor is considerably smaller, roughly on the order of $10^{-5}$. Specifically, in mode (2), the ratio $|\mathcal{P}_w^{QED, (2)}/I-1| \approx$  $6.0
 \times 10^{-5}$, while in mode (3), it's around $6.1 \times 10^{-5}$.

%Let us imagine an experimental setup composed by a laser that could emit  polarized  beam transversely to an external magnetic field $\mathbf{B}_e$ in the direction $\hat{x_3}$. The photon beam is pointing to the Dirac material  $\Delta l$ located in the plane $x_1 x_3$ (parallel to the magnetic field), Fig. (\ref{Exp_Presiones}). The photon beam exerts the radiation pressure Eq. (\ref{Poynting}) over the sheet. The measurement of the non-linear radiation pressures would depend on $\mathcal{L}^D_{\mathcal{GG}}$ and $\mathcal{L}^D_{\mathcal{FF}}$ for a chosen polarization beam. \textbf{Considering  the radiation pressure produced by a laser beam with  intensity $I\sim\langle E_w^2\rangle $} the radiation pressure detected will be  higher (lower) than the classical radiation of the same laser beam travelling in an empty vacuum.  
%\textbf{One can measure the radiation pressure as the difference between the two modes of polarization,  $\Delta\mathcal{P^D}_w= \mathcal{P^D}_w(0)-  \mathcal{P^D}_w^{(2,3)}(B_e)$ %= \frac{1}{2} (\mathcal{M}^{ph-B_e (2)}-\mathcal{M}^{ph-B_e (3)} )B_e$
%, with which it is only necessary to know the photon-field magnetization of the material, but first it has to be isolated from the total magnetization and measure with good precision as we discussed before, which could be a challenge still.}

Of course, employing a comparable experimental setup allows for the measurement of different phases and phase velocities (birefringence) of the photon beam traversing the sheet compared to one propagating without it, both transversely propagating in the magnetic field. The birefringence is characterized by the difference between $\mathcal{L}^D_{\mathcal{GG}}$ and,  $\mathcal{L}^D_{\mathcal{FF}}$ as was pointed out in Eq. (\ref{birefringence}).

%%%%%%%%%%%%%%%%%%%%%%%%%%%%%%%%%%%%%%%%%%%%%%%%

\section{Conclusions}

In this paper, we extrapolate on one side and extend on the other our previous studies on photon propagation in a constant magnetic field of the QED to the context of 3D Dirac materials.

We start from the expansion up to the second order of the photon fields of the extended Euler-Heisenberg effective non-linear Lagrangian, considering low-energy photons $\omega \ll 2mc^2$ ($\omega \ll 2\Delta$) and replacing the fine-structure constant $\alpha$ by $\alpha_D$, which depends on the Fermi velocity $v_F$ instead of $c$, to describe 3D Dirac materials.

Extended effective Lagrangian to finite temperature and density is used to study the properties of a photon propagating in a magnetized medium by expanding the effective Lagrangian up to the second order in the photon fields, allowing us to study the thermodynamic properties of electrons in the presence of the magnetic field and photon propagating in 3D Dirac materials.
%
%\textcolor{red}{
The first term of the expansion, the zero-order term (at $E_w=B_w=0$), is the effective Lagrangian of Euler-Heisenberg or the thermodynamic potential of electrons. So, we obtained the magnetization of the electron and the virtual pairs interacting with the magnetic field by computing the derivative of $\mathcal{L}_D$ with respect to the magnetic field.
Alternatively,
the first and second-order terms of the expansion describe the properties of the photon interacting with the magnetic field via the medium and virtual pairs (vacuum).
%}

%\textcolor{red}{
The magnetization of a Dirac material (due to electrons) depends on $\alpha_D$ and the energy gap $\Delta$. The magnetic properties have been illustrated for the  Dirac materials: $Bi_{1-x}Sb_{x}$, $Pb_{1-x}Sn_{x}Te$, $Ta_3As_4$, and graphene-like, and the properties are determined by the value of the energy gap of each material.
%}

As we expected mathematically and physically, since the medium magnetization comes from the first term of the expansion in series of effective Lagrangian, the electron magnetization is higher than the photon magnetization.
In addition, the magnetization caused by the interaction of photons with the magnetic field is three or more orders of magnitude lower, depending on the material characteristics. Electron and photon magnetization preserve the paramagnetic behavior.
%in zero temperature limit and high density.
%
We noted that with the increase of the magnetic field, the vacuum magnetization grows, becoming almost constant. Its contribution changes the magnetization by a quantity that could be detected at magnetic field values reachable in the laboratory. This conclusion reinforces the importance of considering the vacuum properties when studying Dirac materials. 

We have obtained the dispersion law, the phase velocity, the refractive index of photons propagating in the Dirac vacuum, electric permittivity, and the magnetic susceptibility that the photon feels.
The results agree with those obtained in \cite{Neves2023uxi}.
We have discussed the energy density, pointing vector, and radiation pressures of the photon propagating transverse to the magnetic field starting from the energy-momentum tensor calculated ``a la Hilbert" for effective non-linear electrodynamics for Dirac material at zero temperature and density. These quantities depend on the polarization mode. We have discussed that in our framework, the pressure target in the direction $\mathbf{x_2}$ does not coincide with the Poynting vector. That is a consequence of the symmetry breaking that the magnetic field produces.

Although our model for Dirac materials is simplified from the viewpoint of material science, it has all the wealth of quantum theories. It could be used to study other phenomena that appear in QED and their analogies in Dirac materials.
On the other hand, the photon vacuum properties were studied for arbitrary magnetic field values, reproducing the weak and strong magnetic field limits. 
Besides, our study could be extended to other non-linear Lagrangians that contribute to extracting some general properties of Dirac materials.

Dirac materials present a promising avenue for probing QED properties,
%In comparison to the modest correction factor of $10^{-5}$ to $10^{-4}$  observed in traditional QED scenarios, 
since non-linear corrections in magnetization, radiation pressure, and birefringence are amplified up to $10^3$ times QED corrections. % This significant increase underscores the heightened sensitivity and potential for experimental validation of QED phenomena within Dirac materials. 
 Dirac materials may be a medium where the vacuum properties of QED may be tested.

\section{Acknowledgments}
The authors thank A. to P\'erez Garc\'ia for her contribution to our previous work. A.R.J, A.P.M, and E.R.Q. were supported by the project of No. NA211LH500-002 from AENTA-CITMA, Cuba. A.R.J. acknowledges the financial support provided by the  Giersch Foundation.

%\bibliographystyle{unsrt}
%\bibliography{references_Dirac.bib}

\appendix

\section{Effective Lagrangian of QED at constant magnetic field from imaginary time formalism to proper time}\label{equivalencia}

Let us show the fundamental tricks to get the equivalent representation of  $\mathcal{L}_{eff} (B_e)$ and $\mathcal{L}_{eff} (B_e,\mu,T)$  from one loop QED in presence of a magnetic field in imaginary time formalism to the Schwinger's proper method. We also show how to regularize the ultraviolet divergence of $\mathcal{L}_{eff} (B_e)$. 
At a constant magnetic field, the propagator of an electron in momentum space has the form
\begin{equation}\label{Green-F-L}
G^{-1}_n(\overline{p})= \overline{p}\cdot\gamma-m,
\end{equation}
with the notation ${\overline{p}}=(ip^{4},0, \sqrt{2eB_e n},p^{3})$ over the Landau numbers $n=0,1,2,...$ in Euclidean space, $\gamma$ are the Gamma-matrices.
In the momentum space, the effective Lagrangian has the form \cite{JorgitoyHugo, AuroraPRD2015}

\begin{equation}\label{Grand-Potential-4}
\Omega(B_e,\mu, T)= -\frac{eB_e}{\beta}\left[
\sum_{p_4}\int\limits_{-\infty}^{\infty}\ln \det G^{-1}_0(\overline{p}^*)+
\sum_{\sigma=\pm1}\sum_{n=0}^{\infty}\sum_{p_4}\int\limits_{-\infty}^{\infty}\frac{dp_3}{(2\pi)^2} \ln \det G^{-1}_n(\overline{p}^*)\right],
\end{equation}
where $\sigma=\pm 1$ are the eigenvalues of the spin, $p_4=i\omega_m$, $\omega_m$ is the Matsubara frequencies $\omega_m=k_B T(2m+1)\pi$ with $m=0,1,2...$ and, $p_3$ is the momentum in $\hat{x_3}$ -direction.

Performing the sum over Matsubara frequencies and calculating the determinants in Eq. (\ref{Grand-Potential-4}), we obtain

\begin{equation}
 \mathcal{L}_{eff}(B_e,\mu,T)=
 -\frac{1}{2} \frac{eB_e}{4\pi^2} \int_{-\infty}^{\infty}  dp_3  \sum_{\sigma l} \vert E_{\sigma n}\vert + \frac{eB_e}{4\pi^2} \int_{-\infty}^{\infty}  dp_3  \sum_{\sigma ,n} \frac{1}{\beta} \ln(1+e^{-\beta\vert E_{\sigma n}  -\mu\vert})(1+e^{-\beta\vert E_{\sigma n} +\mu\vert})\label{Termog},
 \end{equation}
 \noindent  where $E_{\sigma n}=\sqrt{p_3^2+m^2+eB_e(2n+1+\sigma)}$, the first term of Eq. (\ref{Termog}) corresponds to the $\mathcal{L}_{eff}(B_e)$ the second $\mathcal{L}_{eff}(B_e,\mu,T)$. Consider first $\mathcal{L}_{eff}(B_e)$
\begin{equation}
\mathcal{L}_{eff}(B_e)= \frac{eB_e}{4\pi^2} \int_{-\infty}^{\infty}  dp_3  \sum_{\sigma ,n}\vert E_{\sigma n} \vert,\label{vacBimaginarytime2}
\end{equation}
using the integral representation for  $E$ 
\begin{equation}
E=\int ds (1-e^{-E^2s}) s^{-3/2}.
\end{equation}
we get 
\begin{equation}
\mathcal{L}_{eff}(B_e)=-\frac{eB_e}{8\pi^2} \sum_{n=1}^{\infty}\sum_{\sigma=\pm 1}\int_{\epsilon}^{\infty} \frac{ds}{s^2}\left(2e^{-(m^2+eB_e(2n+\sigma+1))s} \right),
\end{equation}
%
%\begin{equation}
%=-\frac{1}%{8\pi^2}\int_{\epsilon}^{\infty} %\frac{ds}{s^3}\left(2e^{-m^2s} %((eB_es)coth (eB_es)-1-%\frac{(eB_es)^2}%{3}\right)\label{vacBpropertime}.
%\end{equation}
by performing the sum over the spin $\sigma$, we obtain the integral
\begin{equation}
\mathcal{L}_{eff}(B_e)=\frac{eB_e}{8\pi^2}\sum_{n=1}^{\infty}\int_{\epsilon}^{\infty} \frac{ds}{s^2}\left(2e^{-(m^2+2eB_en)s}-1\right).
\end{equation}

Let’s now do the sum over Landau levels using
\begin{equation}
\sum_{n=1}^{\infty} x^n=\frac{1}{1-x},
\end{equation}
$\mathcal{L}_{eff}(B_e)$ yields
\begin{equation}
\mathcal{L}_{eff}(B_e)=\frac{eB}{8\pi^2}\int_{\epsilon}^{\infty} ds\frac{e^{-m^2s}}{s^2}\left( \frac{1}{1-e^{-2eB_es}}-1 \right),
\end{equation}
in terms of hyperbolic trigonometric function, we have
\begin{equation}
\mathcal{L}_{eff}(B_e)=\frac{eB_e}{8\pi^2}\int_{\epsilon}^{\infty} \frac{ds}{s^2}\left(e^{-m^2 s} \coth(2eB_es)\right),
\end{equation}
\noindent where $\coth(2eB_e s)$ causes that the integral has an ultraviolet divergence.
Expanding
$\coth(2eBs)$ in powers of small $s$ we obtain
\begin{equation}
\coth(2e B_e s) \sim  \left (1+\frac{(eB_es)^2}{3}\right). \label{renor}
\end{equation}
The prescription to regularize the integral consists of adding and subtracting divergencies terms Eq. (\ref{renor})  obtaining,

\begin{equation}
\mathcal{L}^R_{eff}(B_e)=-\frac{1}{8\pi^2}\int_{\epsilon}^{\infty} \frac{ds}{s^3}\left(2e^{-m^2s} ((eB_es)\coth (eB_es)-1-\frac{(eB_es)^2}{3}\right).\label{vacB}
\end{equation}
To archive the effective Lagrangian for the vacuum of QED  in the presence of a constant electric field is obtained from Eq. (\ref{vacB}) substituting $B_e=iE_e$, 
\begin{equation}
\mathcal{L}^R_{eff} (E_e)=-\frac{i}{8\pi^2}\int_{\epsilon}^{\infty} \frac{ds}{s^3}\left(2e^{-m^2s} ((eE_es)\cot(eE_es)-1-\frac{(eE_es)^2}{3}\right).\label{{vacE}}
\end{equation}
The regularized effective Lagrangian Eq. (\ref{eqEHTfinita}) dependent on the electromagnetic fields can be  straightforwardly obtained as  $\mathcal{L}^R_{eff}$ \cite{Acatrinei:2019rwo}, 
\begin{align}	&\mathcal{L}^R_{eff}(\tilde{a},\tilde{b})=-\mathcal{F}-\frac{1}{8 \pi^2}\int_{0}^{i\infty} \frac{d s}{s^{3}} e^{-i(m^2-i\epsilon) s } \nonumber\\
	&\times\left [ (e s)^{2} \tilde{a} \tilde{b} \coth (e \tilde{a} s) \cot(e \tilde{b} s)-\frac{(es)^2}{3}(\tilde{a}^2-\tilde{b}^2)-1
	\right ],
 \label{eqEHTfinita2}
\end{align}

%\subsection{effective Lagrangian of %QED in presence of magnetic %field, at finite T and $\mu$}
Let us consider the second term of the effective Lagrangian Eq. (\ref{Termog}) dependent on temperature and density, and we check it is equivalent to Eq. (14)
\begin{align}
\mathcal{L}_{eff}(B_e,\mu,T) &= %\frac{eB}{4\pi^2} \sum_{\sigma %,n} \frac{a_n}{\beta} \int_{-%\infty}^{\infty} dp_3 \ln(1+e^{-%\beta|E_{\sigma n} -\mu|})(1+e^{-%\beta|E_{\sigma n} +\mu|})\\
\frac{eB_e}{4\pi^2}\sum_{\sigma ,n} \frac{a_n}{\beta}\int_{-\infty}^{\infty} dp_3 \left (\ln (1+e^{-\beta|E_{\sigma n} -\mu|}) +\ln(1+e^{-\beta|E_{\sigma n} +\mu|})\right ), 
\end{align}

We first expand the logarithm  in series
$\ln(1+ x)=\sum_{k=1}^{\infty} \frac{(-1)^{k-1}}{k}x^k$,
and using the integral representation of $e^{-\beta k \sqrt{p_3^2 + m^2_n}}$  
\begin{align}
&=- \frac{eB_e}{4\pi^2}\sum_{n} \frac{a_n}{\beta} \sum^{\infty}_{k=1} \frac{(-1)^{k}}{k}\cosh(\mu \beta k)\int_{-\infty}^{\infty}dp_3\underbrace{e^{-\beta k \sqrt{p_3^2 + m_n^2}}}_{=\frac{\beta k}{2} \int_{-\infty}^{\infty}\frac{dt}{\sqrt{\pi} t^3}e^{-\frac{\beta k}{4t}-st}}, \\
&=-\frac{eB_e}{4\pi^2}\sum_{n} a_n \sum^{\infty}_{k=1}  (-1)^{k}\cosh(\mu \beta k)\int_{-\infty}^{\infty} dp_3 \int_{0}^{\infty} \frac{dt}{\sqrt{\pi} t^{3/2}}e^{-\frac{\beta}{4t}-(p_3^2 + m_n^2)t)}, \\
&=-\frac{eB_e}{4\pi^2} \sum_{n} a_n \sum^{\infty}_{k=1}  (-1)^{k}\cosh(\mu\beta k)\int_{-\infty}^{\infty} \frac{dt}{\sqrt{\pi}t^{3/2}}e^{-\frac{\beta^2k^2}{4t}-m_n^2 t} \left(\frac{-\beta k}{2}\right)\underbrace{\int_{0}^{\infty} dp_3 e^{-t p_3^2}}_{\sqrt{\pi/t}}, \label{medBimaginarytime}\\
&=\frac{eB_e}{8\pi^2} \sum^{\infty}_{k=1}  (-1)^{k}\cosh(\mu\beta k)\int_{0}^{\infty} \frac{dt}{t^{2}}e^{-\frac{\beta^2k^2}{4t}-m_n^2 t} \underbrace{\sum_{n=0}^\infty a_n e^{-2 e B_e n t}}_{\coth{(eB_e t)}},\\
&=\frac{eB_e}{8\pi^2} \sum^{\infty}_{k=1}  (-1)^{k}\cosh(\mu\beta k)\int_{0}^{\infty} \frac{dt}{t^{2}}e^{-\frac{\beta^2k^2}{4t}-m_n^2 t} \coth{(eB_e t)}
\end{align}

\subsection{Lagrangian derivatives at $T=\mu=0$}\label{LagrangianderivativesatT0mu0}

At $T=\mu=0$ the integrals of Lagrangian derivatives $\mathcal{L}^f_{\mathcal{F},\mathcal{FF}, \mathcal{GG}}$ are reduced to the previously calculated in \cite{PerezGarcia2022kvz, DittrichLibro}, after perform the integration  their explicit form remain

\begin{align}
\mathcal{L}^D_{\mathcal{F}}=\frac{\partial \mathcal{L}}{\partial \mathcal{F}}\bigg|_{f=0}&=%-\mu{0}^{-1}
\frac{\alpha_D}{2 \pi \mu_{0}}\left (-\frac{1}{3}+2 h_D^{2}+8 \zeta^{\prime}(-1, h_D)-4 h_D \ln \Gamma(h_D)+2 h_D \ln h_D-\frac{2}{3} \ln h_D+2 h_D \ln 2 \pi\right ),\nonumber \\
\mathcal{L}^D_{\mathcal{FF}}=\frac{\partial^2 \mathcal{L}}{\partial^2 \mathcal{F}}\bigg|_{f=0}&=\frac{\alpha_D}{2\pi \mu_{0}B^2}\left (\frac{2}{3} +
4 h_D^2 \psi(1+h_D)-2h_D-4h_D^2-4h_D \ln \Gamma(h_D) +2h_D \ln 2\pi -2h_D\ln h_D\right ), \nonumber \\
\mathcal{L}^D_{\mathcal{GG}}=\frac{\partial^2 \mathcal{L}}{\partial^2 \mathcal{G}}\bigg|_{f=0}&=\frac{\alpha_D}{2\pi\mu_{0}B^2}\left( -\frac{1}{3}-\frac{2}{3}\left( \psi(1+h_D) -2h_D^2 +(3h_D)^{-1}\right ) +8\zeta^{\prime} \right).\label{integralsvac}
\end{align}

\noindent where $h_D= \frac{\Delta^2}{eB_e}$. 
Similar results of the integrals Eq.(\ref{integralsvac}) are obtained for a pure background electric field and for the background of an orthogonal electric and magnetic field,  doing $h_D\rightarrow \frac{\Delta^2}{2e\sqrt{\mathcal{F}}}$. 
%{\it{revisar y hacer compatibles estas expresiones con las de $T,\mu=0$ (vacío), $+def. a,b$ empleadas, $s \rightarrow -it$}}.

\subsection{Derivatives Lagrangian $T=0, \mu\neq 0$}\label{DerivativesLagrangianT0mune0}
Let's compute the integrals for $\mathcal{L}_{\mathcal{F},\mathcal{FF},\mathcal{GG}}^{(\mu,T),D}$. To achieve this, we take advantage of the equivalence in one-loop approximation between the effective Euler-Heisenberg (EH) Lagrangian in the infrared regime (as presented in this paper)  %Quantum Electrodynamics (QED) in a one-loop approximation, 
and the off-shell photon's polarization operator  \cite{Shabad:2011hf}. 

The equivalence between both procedures is given by relations between the coefficients of the quadratic effective EH Lagrangian ($\mathcal {L}_{\mathcal{F},\mathcal{FF},\mathcal{GG}}^{(\mu,T),D}$) and the eigenvalues of the polarization operator (see details in \cite{Shabad:2011hf} Eq. 21).
This equivalence may be extended to the temperature and density correction term of the EH-effective Lagrangian. 
The relationship between the eigenvalues of the polarization operator $\Pi_{\mu\nu}$ an the Lagrangian derivatives is the following 
\begin{align}
2 \mathcal{F} \mathcal{L}_{\mathcal{F}}^{(\mu,T),D}=\kappa_{1\mid_{k\rightarrow 0}}, \\
2 \mathcal{F}\mathcal{L}_{\mathcal{FF}}^{(\mu,T),D}=(\kappa_{1}-\kappa_{3})_{\mid_{k\rightarrow 0}}  , \\
2 \mathcal{F}\mathcal{L}_{\mathcal{GG}}^{(\mu,T),D}= (\kappa_{1}-\kappa_{2})_{\mid_{k\rightarrow 0}} .
\end{align}
Then, we start from the eigenvalues of the polarization operator at finite temperature and density $\kappa_{1},\kappa_{2},\kappa_{3}$  obtained  in  \cite{Hugo1}, in the degenerate limit (zero temperature and finite density), obtaining
\begin{align}
\mathcal{L}^{(\mu),D}_{\mathcal{F}}(\mu,B_e)&=-\frac{\alpha_D}{2\pi B_e B_c}\sum_{n=0}^{n_{\mu}} a_n
\int \frac{dp_3\theta(\mu-E_{\sigma n})}{\sqrt{p_3^2 v_F^2+\Delta_n^2}}\nonumber\\
&=-\frac{\alpha_D}{2\pi B_e B_c v_F}\sum_{n=0}^{n_{\mu}} a_n ln\left(1+\frac{2p_F v_F(\mu+p_F v_F)}{\Delta_n }\right ),\\
{\mathcal L}^{(\mu),D}_{\mathcal{FF}}(\mu,B_e)
%&=\frac{\alpha_D}{2\pi B_e B_c}\sum_{n=0}^{n_{\mu}} \alpha_n \int \frac{dp_3\theta(\mu-E_{\sigma n})}{2\sqrt{p_3^2+\Delta_n^2}}\nonumber\\%
%&=\frac{\alpha_D}{2\pi B_e B_c}\sum_{n=0}^{n_{\mu}} \frac{\alpha_n}{2} ln\left(1+\frac{2p_F(\mu+p_F)}{2\Delta_n }\right )  \label{LGGLFFM},\\
&=\frac{\mathcal{L}^{\mu}_{\mathcal{F}}}{2},\\
{\mathcal L}^{(\mu),D}_{\mathcal{GG}}(\mu,B_e)&=-\frac{\alpha_D}{4\pi B_e B_c }\sum_{n=0}^{n_{\mu}} a_n \left( \int \frac{dp_3\theta(\mu-E_{\sigma n})}{\sqrt{p_3^2 v_F^2+\Delta_n^2}}-4\int \frac{dp_3\theta(\mu-E_{\sigma n})}{E_{\sigma n}^3} \right) ,\nonumber\\
&=-\frac{\alpha_D}{4\pi B_e B_c v_F}\sum_{n=0}^{n_{\mu}} a_n \left (ln\left(1+\frac{2p_F v_F(\mu+p_F v_F)}{\Delta_n^2 }\right )-4\frac{p_F}{\mu}\right ).\label{LGGLFFM2}
\end{align}

\end{document}